\documentclass[preprint,aps,amsmath,amsfonts,amssymb,superscriptaddress,showpacs]{revtex4}
\usepackage{epsfig}
\usepackage{graphics}
\usepackage{color} 
\usepackage{tikz} 
\usepackage{fancybox}
\usepackage{framed}

\begin{document}
\title{Anti-Coarsening and Complex Dynamics of Step Bunches on Vicinal Surfaces during Sublimation}
\author{Marian Ivanov}
\affiliation{Institut f\"ur Theoretische Physik, Universit\"at zu K\"oln, 50937 K\"oln, Germany}
\author{Vladislav Popkov}
\affiliation{Interdisziplin\"ares Zentrum f\"ur komplexe Systeme, Rheinische Friedrich-Wilhelms-Universit\"at Bonn,
53117 Bonn, Germany}
\affiliation{Dipartimento di Fisica Teorica "E. R. Caianiello", Universit\`{a} degli Studi di Salerno, Salerno,
Italy}
\author{Joachim Krug}
\affiliation{Institut f\"ur Theoretische Physik, Universit\"at zu K\"oln, 50937 K\"oln, Germany}

\begin{abstract}
A sublimating vicinal crystal surface can undergo a step bunching
instability when the attachment-detachment kinetics is
asymmetric, in the sense of a normal Ehrlich-Schwoebel effect. Here we
investigate this instability in a model that takes into account the
subtle interplay between sublimation and step-step interactions, which
breaks the volume-conserving character of the dynamics assumed in
previous work. On the basis of a systematically derived continuum
equation for the surface profile, we argue that the
non-conservative terms pose a limitation on the size of emerging step
bunches. This conclusion is supported by extensive simulations of the
discrete step dynamics, which show breakup of large bunches into
smaller ones as well as arrested coarsening and periodic oscillations
between states with different numbers of bunches.  
\end{abstract}
\date{\today}
\maketitle
\section{\label{Sec:Intro}Introduction}
We consider a vicinal crystal surface of parallel oriented steps in contact with the vapor phase of the same material. 
By varying the sample temperature the crystal can start to sublimate
or grow in the step flow mode 
\cite{Jeong1999,Krug2005,PierreLouis2005,Krug2010}. 
In the presence of an asymmetry in the kinetics at the steps bordering
a terrace, perturbations in the step flow may grow such that the
step profile undergoes a \textit{step bunching} instability. There are two questions of general interest. First, what are the physical conditions required for such an instability? And second, how does the surface morphology evolve once the instability has appeared?

In recent years many groups have addressed these questions within the
framework of the standard model first introduced by Burton, Cabrera
and Frank \cite{Burton1951} (BCF).
In the absence of additional effects such as electromigration \cite{Stoyanov1991,Sato1999} or strain \cite{Tersoff1995}, 
the basic stability scenario is simple: Preferential attachment to ascending steps
(a normal Ehrlich-Schwoebel effect) leads to step bunching during sublimation, while preferential attachment to descending steps 
(an inverse Ehrlich-Schwoebel effect) implies step bunching during growth \cite{Schwoebel1969,Pimpinelli1994,Uwaha1995,Sato2001,Xie2002,Slanina2005}.  
The symmetry between growth and sublimation suggested by this result is however not complete \cite{PierreLouis2003,Fok2007}:
Whereas the deposition flux in a growth experiment is an externally controlled parameter that is independent of the surface
morphology, the sublimation flux is driven by the surface free energy and hence depends on the curvature of the surface. 
As a consequence, the time evolution of the surface profile conserves the
volume of the film (apart from a constant rate of increase) in the
case of growth but not in the case of sublimation \cite{Krug1997}. 

In the present paper we explore the consequences of the
non-conservative nature of the sublimation dynamics for the linear and nonlinear evolution 
of step bunches. We base our treatment on the 
BCF model in the quasistatic limit including
sublimation, the Ehrlich-Schwoebel effect and step-step interactions.
In previous work we have developed a detailed continuum description of
step bunching in this system, in which non-conservative terms
arising from the interplay between sublimation and capillarity were
however neglected \cite{KrugPimpinelli,popkovkrug}.  Here these terms
are explicitly included and their effect is studied both in the
discrete dynamics of individual steps and on the continuum level. 

Although the coefficients of the non-conserved terms are usually small
under physically realistic conditions, they turn out to have dramatic consequences.
Most importantly, whereas step bunches in conserved systems typically
coarsen indefinitely, here we find that coarsening is arrested when a
maximal bunch size has been reached. Conversely, if the system of
steps is started in an initial condition representing a single large
bunch, anti-coarsening involving the breakup of the initial bunch into
several small bunches is observed. More complex scenarios in which the
number of bunches in the system varies periodically in
time are also possible. 

The article is organized as follows. In the next section we briefly
introduce the BCF-model which forms the starting point of our work. In
Section \ref{Sec:Discrete} the discrete equations of motion for the
steps are presented and the results of a linear stability analysis are
described. In Section \ref{Sec:Cont} we derive a continuum evolution
equation along the lines of \cite{KrugPimpinelli,JoachimKim} and provide a
partial analysis which suggests the existence of an upper bound on the size of step
bunches in the non-conserved case. Section \ref{Sec:Numerics} is
devoted to the numerical exploration of the various scenarios of
nonlinear evolution in the discrete step dynamics, and some
conclusions are presented in Section \ref{Sec:Summary}. Details of the analytic
calculations are collected in the Appendices.

\section{\label{Sec:Model}Model}

\begin{figure}
\centerline{\includegraphics[width=8cm]{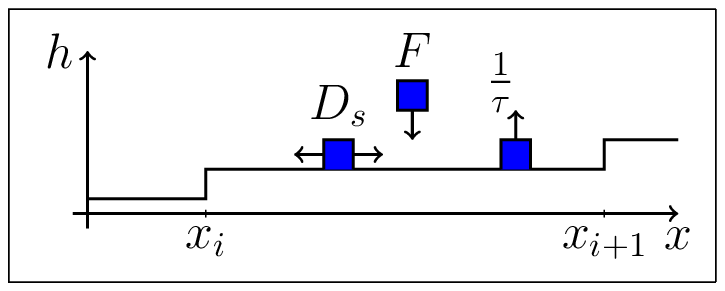}}
\caption{(Color online) Sketch of the elementary processes in the
  Burton-Cabrera-Frank model}
\label{Fig_BCF}
\end{figure}

 On the mesoscopic scale the surface can be reduced to a
 one-dimensional train of steps. The BCF model is based on a
 non-conserved diffusion equation for the concentration profile
 $n_i(x,t)$ on the $i$'th terrace (see Fig.\ref{Fig_BCF}), which reads 
\begin{equation} \label{eq:BCFmodel}
\frac{\partial n_i(x,t)}{\partial t}=D_s\frac{\partial^2 n_i(x,t)}{\partial x^2}-\frac{n_i(x,t)}{\tau}+F.
\end{equation}
The terms on the right hand side correspond to the three processes
sketched in Fig. \ref{Fig_BCF}. The first one is a diffusion term with surface diffusion coefficient $D_s$,
the second term includes the losses of adatoms
during sublimation with desorption rate $1/\tau$, and the
last one describes the gain of adatoms from the surrounding gas phase
with deposition rate $F$. Together diffusion and desorption give rise
to the diffusion length $l_D = \sqrt{D_s \tau}$, defined as the
distance an adatom travels before it desorbs (in the absence of other processes).

Solving \eqref{eq:BCFmodel} in the quasistatic limit $\partial
n_i/\partial t = 0$ one can find the general solution. 
The constants of integration are specified through the following
boundary conditions. We consider a terrace of width $l$ confined
between two steps at the positions $x=\pm l/2$. The condition of mass conservation on both bounding steps defines the system of differential equations  
\begin{eqnarray}\label{eq:bc}\nonumber
D_s\frac{\partial n(x)}{\partial x}=+k_-[n(x)-n_{eq}(x)], \  \  \  \text{at}\   x=-\frac{l}{2},\\
D_s\frac{\partial n(x)}{\partial x}=-k_+[n(x)-n_{eq}(x)], \  \  \  \text{at}\   x=+\frac{l}{2}.
\end{eqnarray}
On the left hand sides the adatom fluxes from the terrace
toward the steps appear, which are compensated by the attachment and
detachment of adatoms that are caused by the difference between
the actual concentration at the step relative to the equilibrium concentrations.

The system \eqref{eq:bc} contains two additional effects.
First, the proportionality constants $k_{-}$ and $k_{+}$ are the
\textit{attachment/detachment kinetic coefficients}, where index the
$+$ ($-$) denotes the coefficient corresponding to
attachment/detachment from below (above) the step. In general the
$k_{\pm}$ are unequal. The case $k_+>k_-$ corresponds to the so
called \textit{Ehrlich-Schwoebel effect} \cite{Schwoebel1969} while in
the converse case we speak about an \textit{inverse Ehrlich-Schwoebel
  effect} \cite{Sato2001,Xie2002,Slanina2005}.  Analogous to the
diffusion length we define the 
\textit{kinetic lengths} $l_{\pm}=D_s/k_{\pm}$ \cite{Krug2005} and
further their dimensionless versions  $l_{\pm}/l_D=l^{\pm}$ (note the
different placement of the indices $\pm$).

The second effect included in the boundary conditions (\ref{eq:bc})
are the (repulsive) \textit{step-step interactions}. 
The expression for the equilibrium concentration is given by the usual
grand canonical formula, which we use in the first order approximation
\begin{equation}
\label{eq:first}
n_{eq}(x_i)\approx n^0_{eq}(1+\mu_i /k_BT).
\end{equation}
The chemical potential $\mu_i$ at the $i$th step depends on the widths
$l_i=x_{i}-x_{i-1}$ of the neighboring terraces according to the law 
\begin{equation}\label{eq:chempot}
\frac{\mu_i}{k_BT} = -g\left [ \frac{l^3}{(l_i)^3} - \frac{l^3}{(l_{i-1})^3}\right ] \equiv g\nu_i,
\end{equation}
which was first derived theoretically for entropic repulsion
\cite{grubermullins}. The ubiquity of repulsive step-step interactions
is well confirmed by experiments, which also show that the dominant
contribution to the amplitude $g$ arises from elastic interactions \cite{Jeong1999,giesen}. 

In order to render $g$ dimensionless we have rescaled the terrace
widths in (\ref{eq:chempot}) by the mean terrace width $l$. 
Note that our definition of $g$ differs from the conventional notation,
where the strength of step interactions is quantified by
the coefficient of the cubic term in the expansion of the surface free
energy in the surface miscut $\theta$, the angle formed by the vicinal
surface
relative to the high symmetry orientation \cite{Jeong1999}. Denoting
the latter coefficient by $\tilde g$, the relation between the two
reads \cite{LiuWeeks}
\begin{equation}
\nonumber
g = \frac{2 \tilde g \Omega}{k_B T} \vert \tan \theta \vert^3,
\end{equation}
where $\Omega$ denotes the atomic area. To give an impression of the
order of magnitude of $g$, for the Si(111) surface at 900$^\textrm{o}$ 
the estimate $\tilde g \approx 0.05$ eV/\AA$^2$ \cite{Jeong1999} yields 
$g \approx 10 \vert \tan \theta \vert^3 \approx 5 \times 10^{-5}$ for
a typical miscut of 1$^\textrm{o}$.

\begin{figure}
\centerline{\includegraphics[width=8cm]{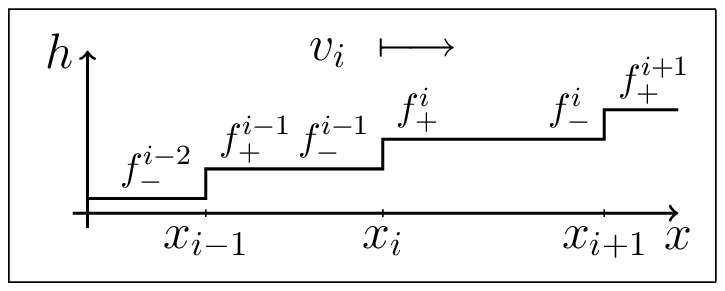}}
\caption{Sketch of the contributions to the velocity of the $i$-th step.}
\label{Fig_velo}
\end{figure}

Together Eqs.(\ref{eq:BCFmodel},\ref{eq:bc},\ref{eq:first},\ref{eq:chempot}) specify the boundary value problem for 
the $n_i(x)$. Having computed the
concentration profiles we can use Fick's first law to find the mass fluxes
from below ($f_+^{i-1}$) and above ($f_-^{i}$) the $i$-th step, which 
sum to give its velocity $v_i=\frac{dx_i}{dt}=f_-^{i}+f_+^{i-1}$ (see Fig. \ref{Fig_velo}).

\section{\label{Sec:Discrete}Discrete view}
\subsection{\label{Sec:stepmotion}Equations of step motion}
In Appendix \ref{app:profile} we derive the concentration profile $n_i(x)$
for the case of sublimation ($F=0$) with step-step repulsion and the
Ehrlich-Schwoebel effect in the quasistatic limit, see
Eq. \eqref{eq:app1a} and  Eq. \eqref{eq:app1e}, which yields
\begin{eqnarray}\label{eq:profile}\nonumber
n_i(x)&=& n^0_{eq} \left [ \frac{(l^+-1)(1+\frac{\mu_i}{kT})e^{-\frac{l_i}{2l_D}}+(l^-+1)(1+\frac{\mu_{i+1}}{kT})e^{\frac{l_i}{2l_D}}}{(l^+l^-+1)\sinh\frac{l_i}{l_D}+(l^-+l^+)\cosh\frac{l_i}{l_D}}\right ]\frac{e^{\frac{x}{l_D}}}{2} +\\
&+& n^0_{eq} \left [\frac{(l^--1)(1+\frac{\mu_i}{kT})e^{-\frac{l_i}{2l_D}}+(l^++1)(1+\frac{\mu_{i+1}}{kT})e^{\frac{l_i}{2l_D}}}{(l^+l^-+1)\sinh\frac{l_i}{l_D}+(l^-+l^+)\cosh\frac{l_i}{l_D}}\right ] \frac{e^{-\frac{x}{l_D}}}{2}.
\end{eqnarray}
From \eqref{eq:profile} we find the following discrete equations of step motion (see Appendix \ref{app:fluxes}) 
\begin{eqnarray}\label{eq:velocity}\nonumber
\frac{dx_i}{dt}&=& \frac{\Omega D_s n^0_{eq}}{l_D}  \left [    \frac{\left (1+\frac{\mu_{i}}{kT}\right )\left (l^-\sinh\frac{l_{i-1}}{l_D}+\cosh\frac{l_{i-1}}{l_D}\right )-\left (1+\frac{\mu_{i-1}}{kT}\right )}{(l^+l^-+1)\sinh\frac{l_{i-1}}{l_D}+(l^-+l^+)\cosh\frac{l_{i-1}}{l_D}}\right ]  + \\
&+&\frac{\Omega D_s n^0_{eq}}{l_D}  \left [  \frac{\left (1+\frac{\mu_{i}}{kT}\right )\left (l^+\sinh\frac{l_{i}}{l_D}+\cosh\frac{l_{i}}{l_D}\right )-\left (1+\frac{\mu_{i+1}}{kT}\right )}{(l^+l^-+1)\sinh\frac{l_{i}}{l_D}+(l^-+l^+)\cosh\frac{l_{i}}{l_D}} \right ].
\end{eqnarray}
The step velocities \eqref{eq:velocity}  include all three length
scales $l_D$, $l_\pm$, $l$ and the functional dependence is quite complicated.
From now on we therefore consider the case of attachment/detachment
limited kinetics ($l_{D} \gg l_{\pm}\gg l $), which is commonly
assumed for the Si(111) surface \cite{LiuWeeks}. In this limit  Eq. \eqref{eq:velocity} can be reduced to the following form (see Appendix \ref{app:limit}):
\begin{eqnarray}\label{eq:discretemodel}
 \frac{dx_i}{dt} = \left (1+g\nu_i\right )R_e \left (  \frac{1+b}{2}l_{i-1}+\frac{1-b}{2}l_{i}\right ) + UR_e(2\nu_i-\nu_{i-1}-\nu_{i+i}),
\end{eqnarray}
with the abbreviations
\begin{eqnarray}\label{eq:parameters}\nonumber
b&=&\frac{k_{+}-k_{-}}{k_-+k_+}\ =\frac{l_{-}-l_{+}}{l_-+l_+},\\ \nonumber
R_e&=&\frac{\Omega n^0_{eq}}{\tau}\, \\ 
U&=&g \tau\frac{ k_-k_+}{k_-+k_+}\ =\frac{g l_D^2}{l_-+l_+}.
\end{eqnarray}
Here $R_e$ represents the constant desorption rate of a homogeneous
step train, where $dx_i/dt=R_el$, $b$ is a
dimensionless measure for the strength of the Ehrlich-Schwoebel effect
and $U$ describes the strength of the relaxation due to the step-step
repulsion. Note that $U$ has the dimension of a length.

In previous work on step bunching during sublimation, where the
instability is induced either by
the Ehrlich-Schwoebel effect or by electromigration, the factor $1+g
\nu_i$ of the first term on the right hand side of (\ref{eq:discretemodel}) was
tacitly replaced by unity \cite{KrugPimpinelli,popkovkrug,LiuWeeks,slava}. The
approximation $1+g\nu_i \approx 1$ may seem plausible because, as we
have seen, $g \ll 1$ under typical experimental conditions. However, it is clear from
the structure of (\ref{eq:discretemodel}) that the presence of this
factor changes the nature of the problem in a qualitative way:  
Considering periodic boundaries for the step train of $M$ steps, so
that $x_{M+1}=x_1+Ml$, and taking the sum of \eqref{eq:discretemodel}
over one period one obtains the configuration-independent constant 
$M l$ on the right hand size only when the term $g \nu_i$ (referred to
in the following as the $g$-term) is neglected. The
full system (\ref{eq:discretemodel}) is fundamentally non-conservative. 

 On the other hand, the step dynamics is exactly conservative for the
 case of pure growth ($F > 0, \tau \rightarrow \infty$) in the same
 limit in which \eqref{eq:discretemodel} was derived. In the case of
 growth we obtain
\begin{eqnarray}\label{eq:growthmodel}
\frac{dx_i}{dt} \approx- F \Omega \left
  (\frac{1+b}{2}l_{i-1}+\frac{1-b}{2}l_{i}\right ) +  \tilde{U} (2\nu_i-\nu_{i-1}-\nu_{i+i})
\end{eqnarray}
which is precisely the conservative version of
(\ref{eq:discretemodel}) with the sublimation rate $R_e$ replaced by
the (negative) growth rate $F \Omega$ and with $\tilde{U}=(g\Omega n^0_{eq}k_-k_+)/(k_-+k_+)$. This implies a basic asymmetry
between sublimation and growth, the consequences of which will be
explored in the following. To this end we will consider $U$ and $g$ in
(\ref{eq:discretemodel}) as independent parameters, in spite of the
proportionality between $U$ and $g$. The conserved model (\ref{eq:growthmodel})
is thus included in (\ref{eq:discretemodel}) as the limiting case $g = 0, U
> 0$.

\subsection{\label{Sec:LSA}Linear stability analysis}
\begin{figure}
\centerline{\includegraphics[width=8cm]{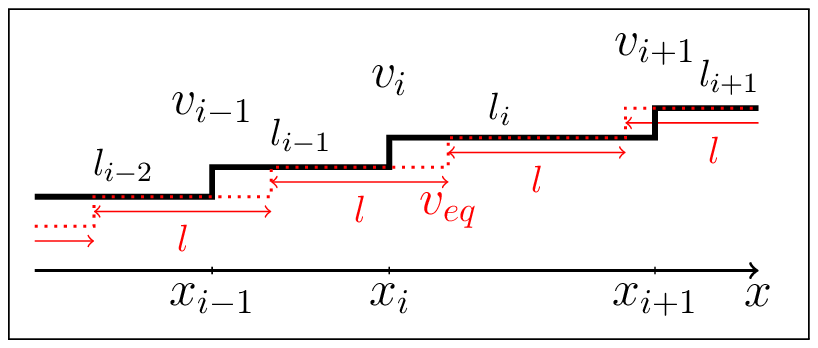}}
\caption{(Color online) Illustration of the linear stability analysis. The dotted
  line shows a homogeneous step train with constant terrace width $l$
  and constant step velocity $v_{eq}$, the full line shows a perturbed step train.}
\label{Fig_velo_eq}
\end{figure}

 The instability form of a step train is step bunching and in this section we look for the linear instability condition as a function of the control parameters. Let us consider the regular situation of equidistant terrace widths $l$ and steps moving with a constant velocity $v_{eq}=f_-(l)+f_+(l)$, see Fig. \ref{Fig_velo_eq}. Now, we disturb the positions of the steps by a small time dependent perturbation $\varepsilon_{n}(t)$:
\begin{eqnarray}\label{eq:LSA}\nonumber
x_n(t)=nl+v_{eq}t+\varepsilon_{n}(t).
\end{eqnarray}
\begin{figure}
\centerline{\includegraphics[width=8cm]{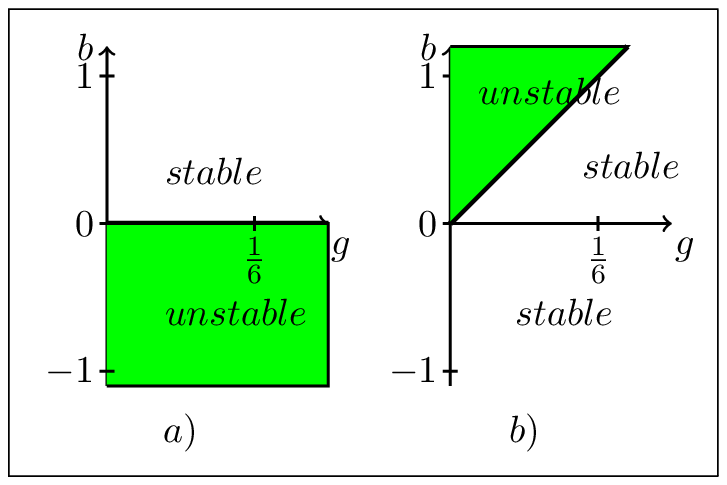}}
\caption{(Color online) Stability diagram in the $(g,b)$-plane for a) growth and  b) sublimation.}
\label{Fig_linstabcond}
\end{figure}
Through substitution of $\varepsilon_{n}(t)$ by the Fourier expression $\varepsilon_0e^{ikn+\omega(k)t}$  we derive a dispersion relation $\omega(k)$.  For instability the real part of $\omega(k)$ has to be positive, i.e. 
\begin{eqnarray}\label{eq:omega}\nonumber
Re[\omega(k)]  \approx A_2k^2+A_4k^4
\end{eqnarray}
with $A_2>0$.
In Appendix \ref{app:linear}, Eq.\eqref{eq:app1r}, we find:
\begin{eqnarray}\label{eq:omega2}\nonumber
Re[\omega(k)]  = \frac{\Omega D_s n^0_{eq}}{l^2_D}\left [ \frac{(k_+-k_-)}{2(k_++k_-)} -3g\right] k^2-3g\frac{\Omega D_s n^0_{eq}}{l}\frac{k_-k_+}{k_-+k_+}k^4.
\end{eqnarray}
For the case of sublimation in the limit of long wavelengths (small $k$) the instability condition is  $b>6g$. This means that the existence of an Ehrlich-Schwoebel effect alone ($b>0$) is not sufficient to cause an instability: Due to the step-step interactions there is a lower limit $6g$ on the strength of the kinetic asymmetry which has to be overcome to cause step bunching during sublimation. 
This effect was previously reported by Fok, Rosales and Margetis \cite{Fok2007} in a more general setting.
On the other hand for the case of growth we have the usual linear instability condition of $b<0$, which corresponds to an inverse Ehrlich-Schwoebel effect of arbitrary strengh.
The resulting asymmetry between sublimation and growth on the level of linear stability analysis is 
illustrated in Fig. \ref{Fig_linstabcond}.

\section{\label{Sec:Cont}Continuum description}

The continuum evolution equation corresponding to the discrete dynamics
\eqref{eq:discretemodel} has the form
\begin{equation}
\frac{\partial h}{\partial t}+\frac{\partial}{\partial x}\left(  -\frac{3g}%
{2}\left(  m^{2}\right)  -\frac{b}{2m}-\frac{1}{6m^{3}}\frac{\partial
m}{\partial x}+\frac{3U}{2m}\frac{\partial^{2}\left(  m^{2}\right)  }{\partial
x^{2}}\right)  =-1+\frac{3gb}{2m}\left(  \frac{\partial m}{\partial x}\right)
^{2} \label{Continuous}%
\end{equation}
where $h(x,t)$ is the surface profile, $m(x,t)=\partial h/\partial x>0$ is the
slope, and primes denote the derivatives with respect to $x$, e.g. $m^{\prime
}=\partial m/\partial x$. Here and below we set the step height and the
average step-step distance $l$ to $1$, and rescale time such that the mean sublimation rate
$R_e = 1$. The details of the derivation can be found
in Appendix \ref{app:cont}. 

For $g=0$, the equation (\ref{Continuous}) reduces to the
one studied in \cite{KrugPimpinelli,popkovkrug}. In that case the transformation to a moving 
frame $h \to h + t$ removes the constant mean sublimation rate on the right hand side and 
gives the equation the form of a conservation law 
\begin{equation}
\label{Conservation}
\frac{\partial h}{\partial t} + \frac{\partial J}{\partial x} = 0,
\end{equation}
with a current $J$ defined by the terms inside the large parentheses on the left hand side
of (\ref{Continuous}). For $g > 0$ we obtain an additional contribution $(3 g/2) m^2$
to the current, and the manifestly non-conservative term $(3 gb/2m) (m')^2$   
appears on the right-hand side of (\ref{Continuous}). In the following we discuss separately
the effects of the conserved and non-conserved contributions that are present when $g > 0$.

\subsection{\label{Sec:stat}Mechanical analog for symmetric stationary bunches}
We first analyze the effect of the conservative term $(3g/2)m^2$ on the left hand side of (\ref{Continuous}).
Along the lines of \cite{KrugPimpinelli}, we consider a stationary solution of (\ref{Conservation}) with 
the current given by 
\begin{eqnarray}
\label{Jfull}
J= -\frac{3g}{2}\left(  m^{2}\right)  -\frac{b}%
{2m}-\frac{m^{\prime}}{6m^{3}}+\frac{3U}{2m}\left(  m^{2}\right)
^{\prime\prime}.
\end{eqnarray}
For a stationary bunch $J \equiv -J_0<0$. Furthermore we neglect the third term on the right hand side of (\ref{Jfull}), 
which breaks the reflection symmetry ($x\rightarrow-x$ and $m\rightarrow-m$). Setting $u=m^2$ ($m>0$) then yields
\begin{eqnarray}\label{fluxy} \nonumber
J_0 =\frac{3g}{2}u+\frac{b}{2}u^{-1/2}-\frac{3Uu^{\prime\prime}}{2}u^{-1/2}
\end{eqnarray}
or
\begin{eqnarray}\label{Newton}
\frac{3U}{2}u^{\prime\prime}=-J_0u^{1/2}+\frac{3g}{2}u^{3/2}+\frac{b}{2} = -\frac{dV_{g}(u)}{du}.
\end{eqnarray} 
Equation \eqref{Newton} can be interpreted as Newton's second law describing a particle coordinate $u$ moving in time $x$.
In this picture a symmetric bunch of width $L$ represents a trajectory of a classical particle moving once back and forth in a potential $V_g(u)$ between the boundary values $u=0$ and $u=u_{max}$ in a time $L$. Up to a constant, the potential $V_g(u)$ can be found through integration of \eqref{Newton},
\begin{eqnarray}\label{Potential}
V_g(u)=\frac{2J_0}{3}u^{3/2}-\frac{3g}{5}u^{5/2}-\frac{b}{2}u=V_0(u)-\frac{3g}{5}u^{5/2}.
\end{eqnarray}
The function \eqref{Potential} contains an additional term compared to
the potential $V_0(u)$ considered in \cite{KrugPimpinelli}. This term
causes a maximum of $V_g(u)$ to appear at some $u^\ast$, whereas
$V_0(u)$ grows monotonically for large $u$. This has the following
consequence. Let us consider a bunch of $M$ steps. For a given set of parameters $b$, $g$ and $U$ this bunch solution corresponds to a particle 
trajectory of total energy $E$ and a certain maximal slope $m_{max}$ with $u_{max} = m_{max}^2 < u^\ast$. Now, let  us increase the bunch width $L$ by adding more and more steps, which increases the energy $E$ and the value of $u_{max}$. As $u_{max} \to u^\ast$ and $E \to V_g(u^\ast)$,
the oscillation period $L$ of the particle diverges, which implies that the maximum slope $m_{max}$ cannot increase beyond 
$\sqrt{u^\ast}$. This is in contrast to the case $g=0$, where the maximum slope scales with the number of steps as $m_{max} \sim M^{2/3}$
and with the bunch width as $m_{max} \sim L^2$ \cite{KrugPimpinelli}. Even on the rather crude level of the stationary bunch approximation
used in \cite{KrugPimpinelli}, the presence of the $g$-term in (\ref{Jfull}) is seen to have a pronounced effect, in 
that it prevents the unbounded steepening of the bunch profile. In the next subsection we shall see that, when bunch motion is taken
into account, the non-conserved nature of the dynamics also prevents the wavelength of the bunch from increasing indefinitely.

\subsection{\label{Sec:fulleq}Moving bunches}

\begin{figure}[ptb]
\centerline{\includegraphics[width=8cm,height=6cm,clip]{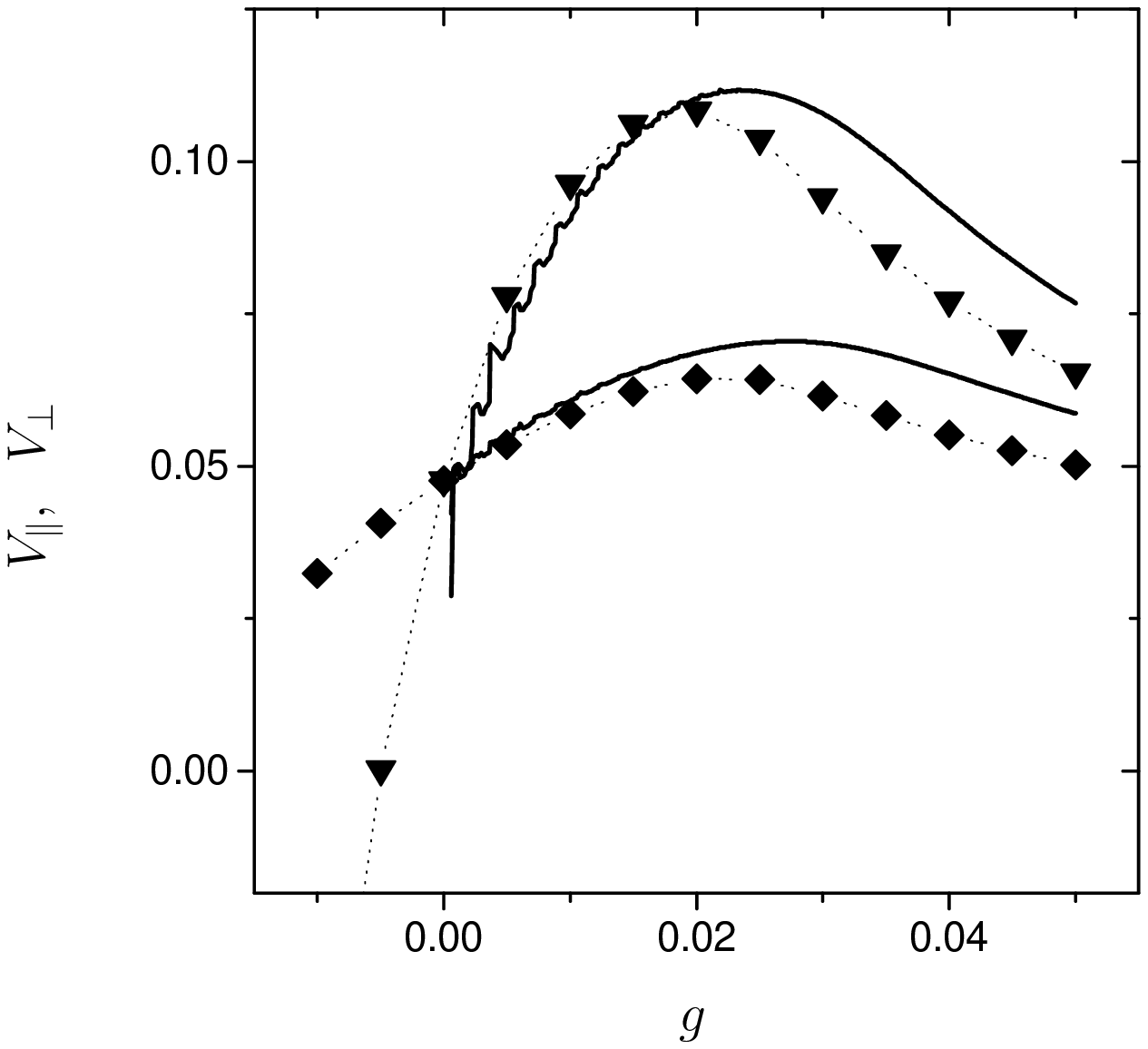}}
\caption{Horizontal
  and vertical bunch speeds $V_\parallel, V_\perp$ 
versus $g$ for the discrete model (diamonds and triangles, respectively), and
for the continuous model (thick lines). Parameters are $b=0.7,U=0.5,M=40$. }%
\label{Fig_M40b7}%
\end{figure}

\begin{figure}[ptb]
\centerline{\includegraphics[width=8cm,height=6cm,clip]{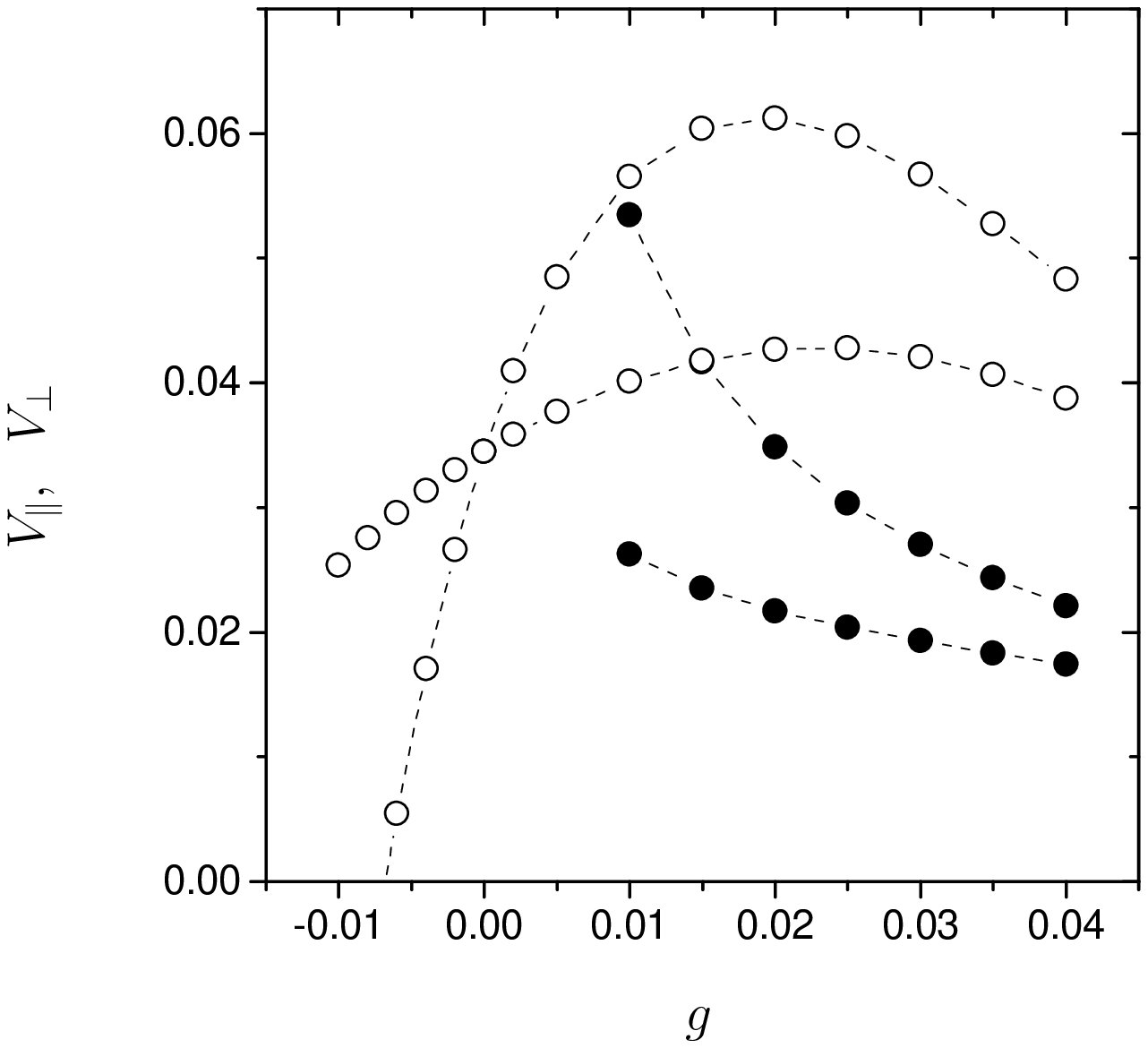}}
\caption{Horizontal
  and vertical bunch speeds as functions of $g$ for the parameters
  $b=0.5,U=0.5$ and $M=40$ (open circles) and $M=80$ (filled circles) obtained from simulations 
of the discrete model. Results for $M=80$ are reported
only for large $g$. In the physical region $g>0$, $V_\perp > V_\parallel$, which
corresponds to step trajectories in the comoving system of coordinates going
downwards (compare to Figs.\ref{Fig_profile} and
\ref{Fig_splitting}). For large $g$, the
absolute values and the difference $V_\perp = V_\parallel$ are
decreasing functions of $M$. 
For small $g$,
$V_\perp - V_\parallel \approx\alpha g$ where $\alpha\approx 3$. From Eq.(\ref{Omega-V})
we get the theoretical estimate $\alpha=$ $\frac{3b^{3}}{8U}M=3.75$}%
\label{Fig_vOmega_b05}%
\end{figure}

In fact step bunches are not stationary, but move both in the horizontal direction (with speed $V_\parallel$) and in the 
vertical direction (with speed $V_\perp$) \cite{popkovkrug}. 
A periodic array of moving bunches is obtained from (\ref{Continuous}) using the travelling wave ansatz
$$h\left(  x,t\right)  =h\left(  x-V_\parallel t\right)  + V_\perp t-t,
$$
which yields the ordinary differential equation
\begin{equation}
- V_\parallel h^{\prime}+V_\perp+\left(  -\frac{3g}{2}\left(  m^{2}\right)  -\frac{b}%
{2m}-\frac{m^{\prime}}{6m^{3}}+\frac{3U}{2m}\left(  m^{2}\right)
^{\prime\prime}\right)  ^{\prime}=\frac{3g b}{2}\frac{\left(  m^{\prime
}\right)  ^{2}}{m}. \label{Model1}%
\end{equation}
Here primes denote derivatives with respect to co-moving space coordinate
$\xi=x-V_\parallel t$ and $m(\xi) = dh/d\xi$. We look for periodic solutions $h(\xi)=h(\xi+M)$. Integrating
(\ref{Model1}) from $0$ till $M$, and dividing by $M$ we obtain
\[
V_\parallel - V_\perp=g\frac{3b}{2M} \int\limits_{0}^{M}\frac{(m^{\prime})^{2}}%
{m}d\xi
\]
which implies $V_\parallel = V_\perp$ in the conserved case \cite{popkovkrug}.
In a linear order approximation in $g$, the integral on the right-hand side
can be estimated by its value at $g=0$. The latter can be computed analytically
using the theory developed in \cite{popkovkrug}, with the result
\[
\int\limits_{0}^{M}\frac{(m^{\prime})^{2}}{m}d\xi=\frac{1}{36Um_{\min}^{2}%
}\approx\frac{(3b)^{2}}{36U}=\frac{b^{2}}{4U}M^{2}.%
\]
Here $m_{min}$ denotes the minimal value of the slope along the profile.
This leads to the prediction
\begin{equation}
V_\perp(g)-V_\parallel(g)=g\frac{3b^{3}}{8U}M\text{ for small }g.\label{Omega-V}%
\end{equation}
This result is in reasonable agreement with the direct measurement of 
$V_\parallel$ and $V_\perp$, performed for the discrete model (Figures \ref{Fig_M40b7} and %
\ref{Fig_vOmega_b05}). From Fig. \ref{Fig_M40b7} we estimate
$V_\perp(g)-V_\parallel(g)\approx\alpha g$ for small $g$ with $\alpha\approx8$,  while
the direct calculation from (\ref{Omega-V}) using the parameters
$b=0.7,U=0.5,M=40$, gives $\frac{3b^{3}}{8U}M=\alpha=10.3$. From Fig.
\ref{Fig_vOmega_b05} we estimate $\alpha\approx3$ \ while the theoretical
prediction gives $\alpha\approx3.75$. 

The result (\ref{Omega-V}) is quite surprising: It tells us that the
difference $V_\perp(g)-V_\parallel(g)$ acquires a singularity at $g = 0$ in the limit
$M\rightarrow\infty$, where continuum theory should become accurate.
To avoid this unphysical singularity, we must assume the existence of
an upper bound on $M$ (and hence, on the wavelength of moving bunch solutions) 
for small but positive values of $g$. Indeed, in the discrete
simulations to be reported in the next section, we will see that the coarsening is arrested at a certain 
maximum wavelength $M_{crit}(g)$. On the other hand, since
in the conservative case ($g=0$) the vertical and horizontal excess speed of the
bunch are equal, $V_\perp = V_\parallel$, we expect from (\ref{Omega-V}) that $\lim
_{g\rightarrow0}gM_{crit}(g)=0$. In other words, we deduce that for small $g$
the critical bunch size $M_{crit}$ must grow more slowly than $1/g$. Note also that
in the conservative case the values of the excess velocities decrease with the bunch
size $M$ as $V_\perp=V_\parallel\approx3b/M$ \cite{popkovkrug}. We observe a similar tendency
(decrease of $V_\perp$ and $V_\parallel$ with increasing $M$) for large enough $g$, 
compare the upper and the lower curves
in Fig.\ref{Fig_vOmega_b05}.

\section{\label{Sec:Numerics}Numerical simulations of the discrete equations }

We simulated the coupled system \eqref{eq:discretemodel} of $M$
non-linear ordinary differential equations of first order with four
independent parameters: $M$ is the number of steps, $g$ is the
parameter describing the non-conserving effects, $b$ is the parameter
describing the Ehrlich-Schwoebel asymmetry and $U$ is the relaxation
parameter due to the step-step interactions. We studied the following
region of the parameter space: $b\in[0,1]$, $U\in[0,1]$, $g\in[0,1]$
and $M<100$. Note that, according to the definition of $U$ in
Eq.(\ref{eq:parameters}), the dimensionless ratio 
$
g l/U = l (l_- + l_+)/l_D^2 
$ should be small compared to unity in order to be consistent with the
assumption $l_D \gg l_\pm \gg l$ under which the model
\eqref{eq:discretemodel} was derived. Since our main interest
here is in exploring the qualitative consequences of the non-conserved
dynamics induced by the $g$-term, rather than in a realistic
description of a particular physical system, we will not always adhere
to this restriction.

 For the numerical integration  we used  an odeint-type procedure
 \cite{Num} with periodic boundary conditions. The height of a single
 step and the average terrace width $l$ are normalized to unity. 
The time scale is normalized by $R_e$ and we measured the time of
integration in time units (t.u.). We started with two qualitatively
different initial conditions: first, an initial 'shock' composed of
densely packed steps and a very large terrace, and second, a randomly
disturbed equidistant step train, where the relative amplitude of
initial fluctuations is chosen to be either small (0.01) or large (0.5). 

\begin{figure}[ptb]
\scalebox{0.60}{\includegraphics{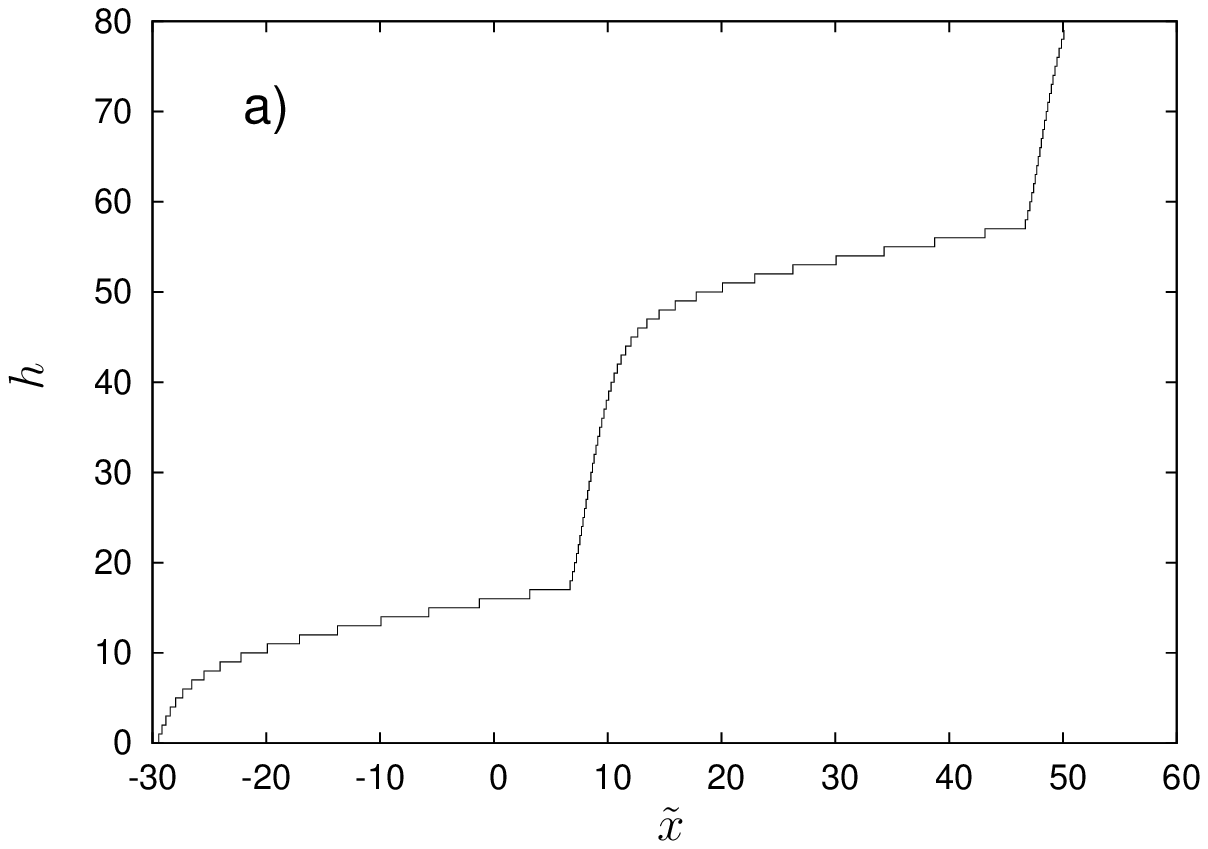}}
\scalebox{0.60}{\includegraphics{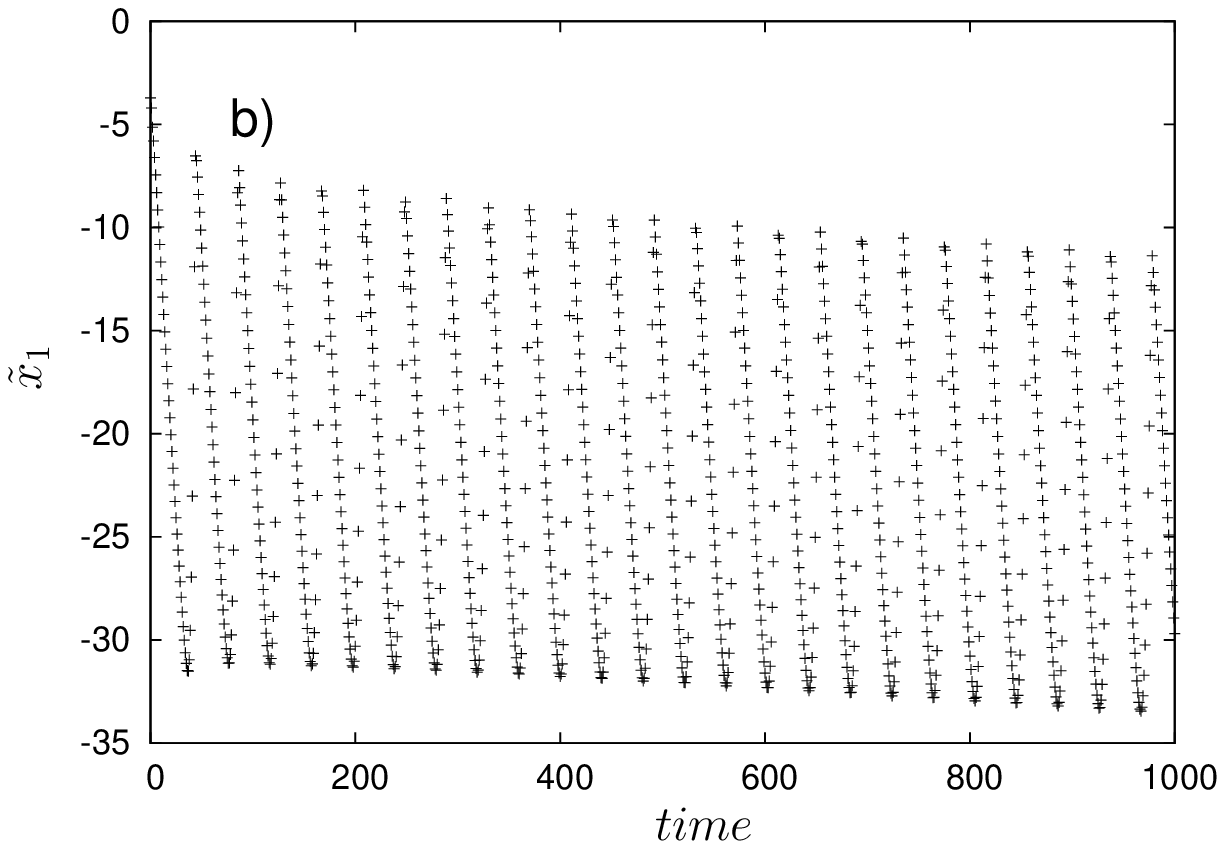}}
\caption{Simulation results for $M=40$, $b=0.1,\  g=0.001,\  U=0.004$,
  starting with an initial shock and periodic boundary conditions. a)
  Typical step train profile after 1000 t.u.; b) Typical time
  evolution of one of the steps in co-moving coordinates.}%
\label{Fig_profile}%
\end{figure}

We define a bunch as a region where the widths $l_i$ of consecutive
terraces are smaller than one. Figure \ref{Fig_profile}a) shows a step
train profile in the linear instability regime ($b>6g$). 
As useful quantitative measures of the bunch geometry we define 
the maximal slope $m_{max} \equiv \max_{i} \{m_i \}$ and the minimal (=maximally
negative) curvature 
$\kappa_{min} \equiv \min_{i} \{\kappa_i \} $, where
\begin{eqnarray}\nonumber 
m_i=\frac{1}{l_i},\  \  \  \  \  \  \  \  \  \  \  \kappa_i=-8\frac{l_{i+1}-l_i}{(l_{i+1}+l_i)^3}.
\end{eqnarray}
  We calculate and plot the positions of the steps in a co-moving
  coordinate system defined by $\tilde{x}_i(t)=x_i(t)-lt$, where $l$
  is the average velocity in the conserved limit ($g = 0$). Because of the additional nonconservative terms, there is a lateral shift after every period, see Fig. \ref{Fig_profile}b) and also Fig. \ref{Fig_splitting}a). 

 For simulations with $b\lesssim6g$ an initial step bunch dissolves
 and approaches the configuration of equidistant steps.
On the other hand for $b>6g$ and increasing $g/U$, the time evolution
of the step train switches unexpectedly to anti-coarsening or arrested
coarsening regimes, depending on the initial conditions. In the
following we describe these two scenarios in more detail.

\subsection{\label{Sec:Anti}Anti-Coarsening}

\begin{figure}
\begin{center}
\scalebox{0.60}{\includegraphics{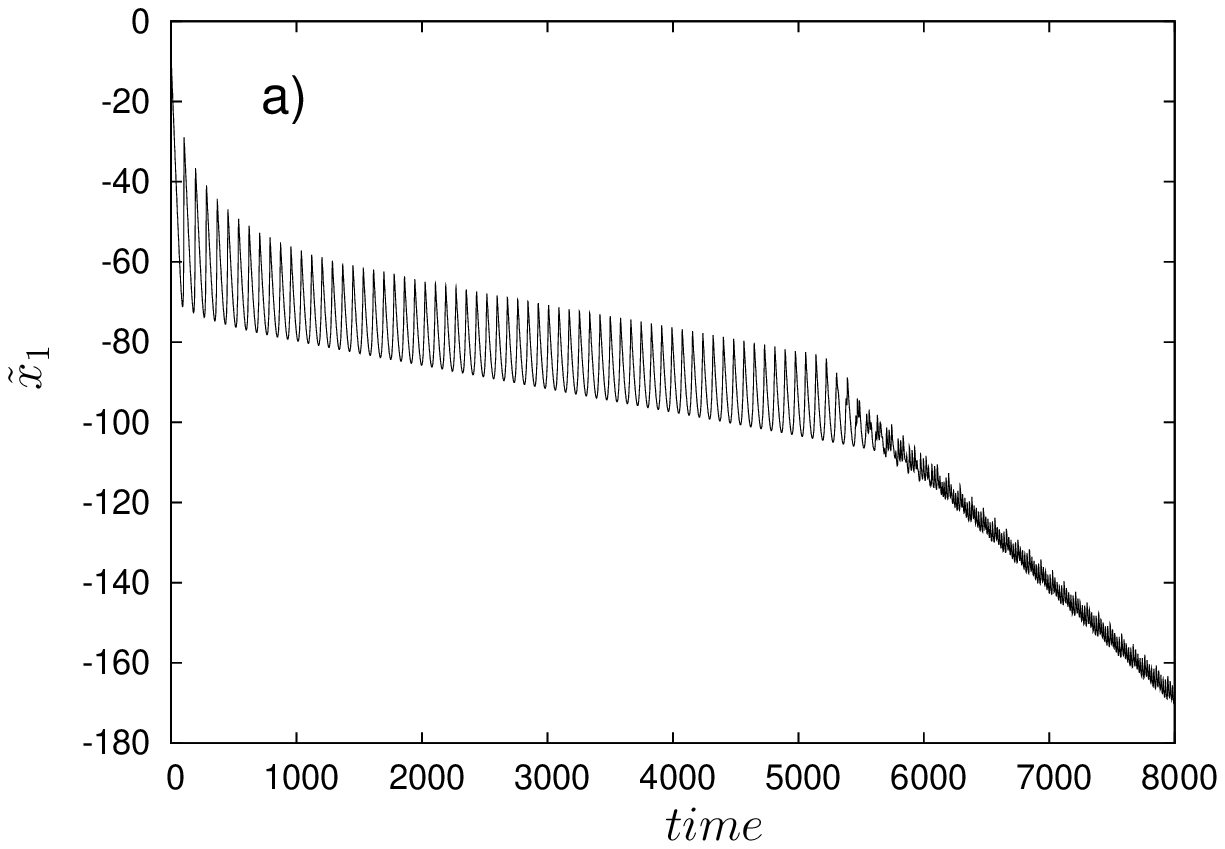}}\  \  \  \  \  \  
\scalebox{0.60}{\includegraphics{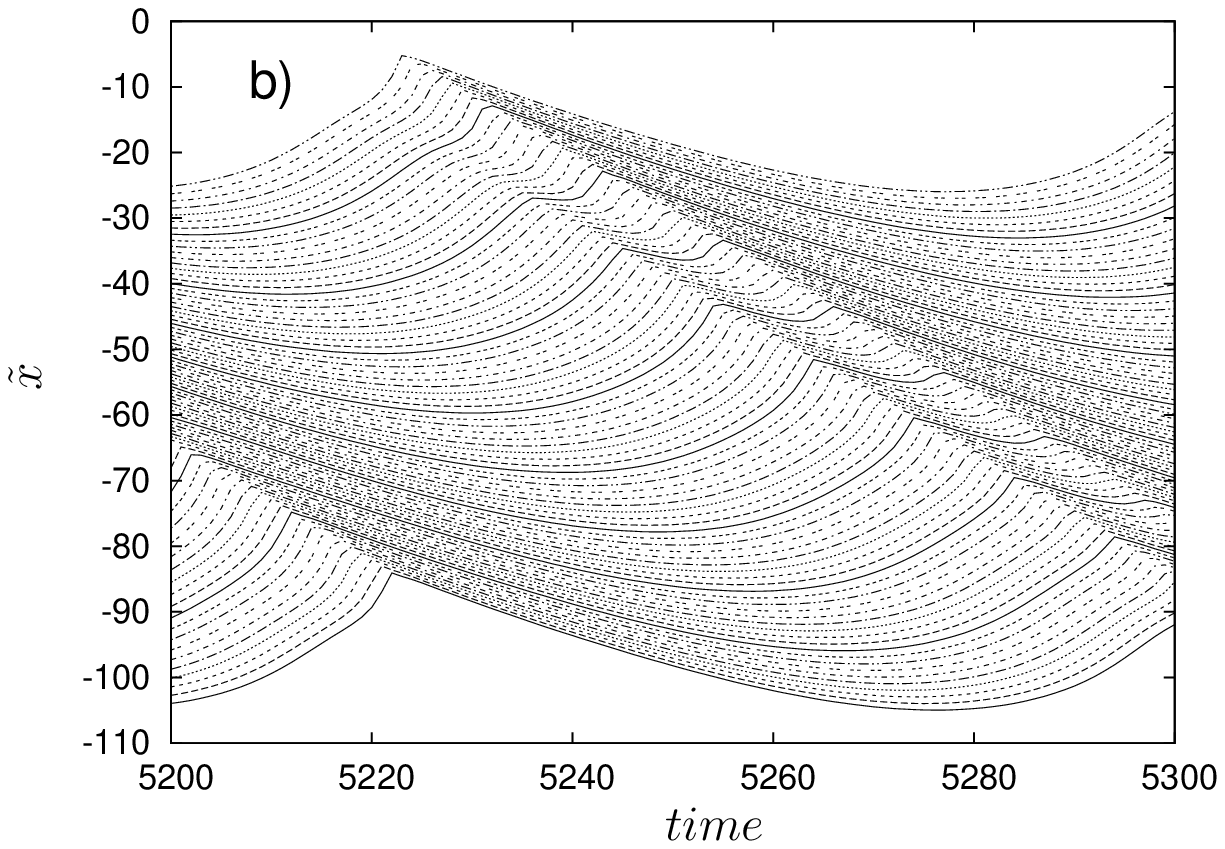}}  \\
\scalebox{0.60}{\includegraphics{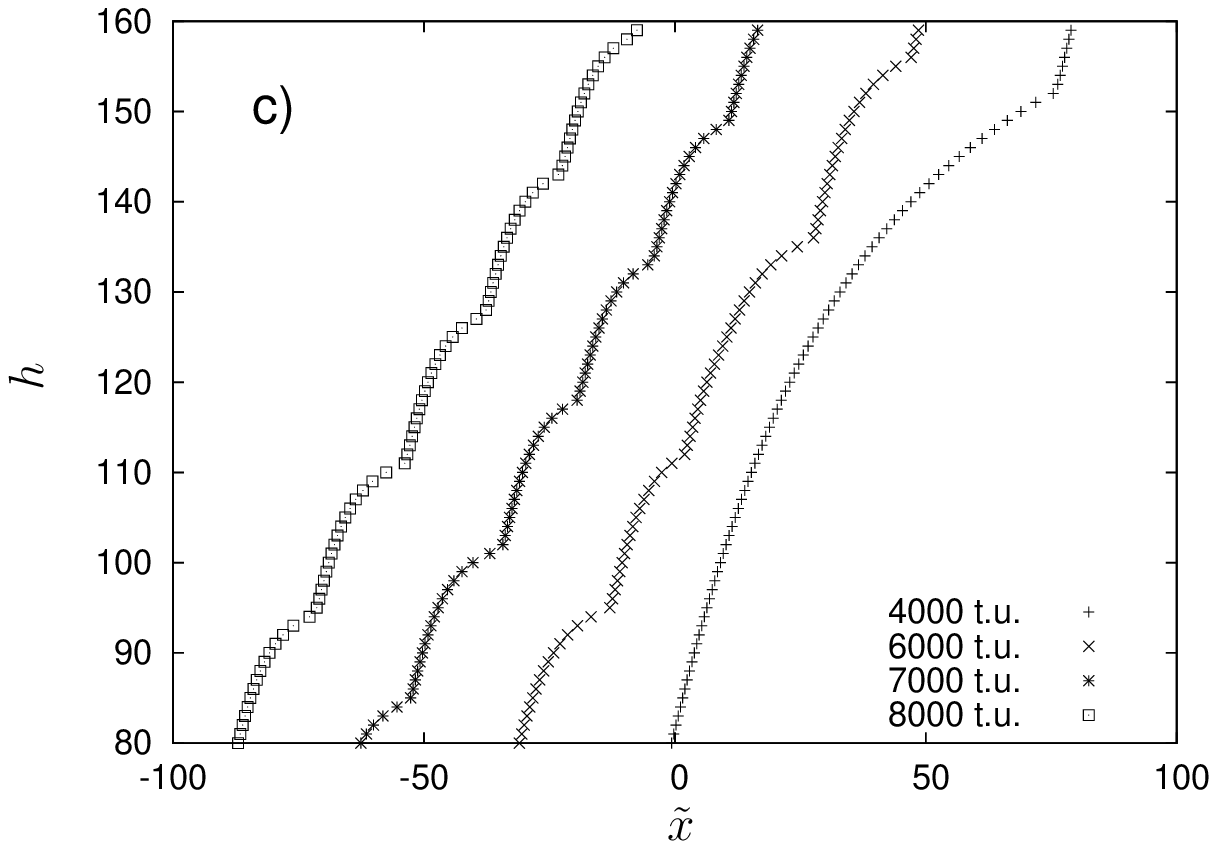}} 
\scalebox{0.60}{\includegraphics{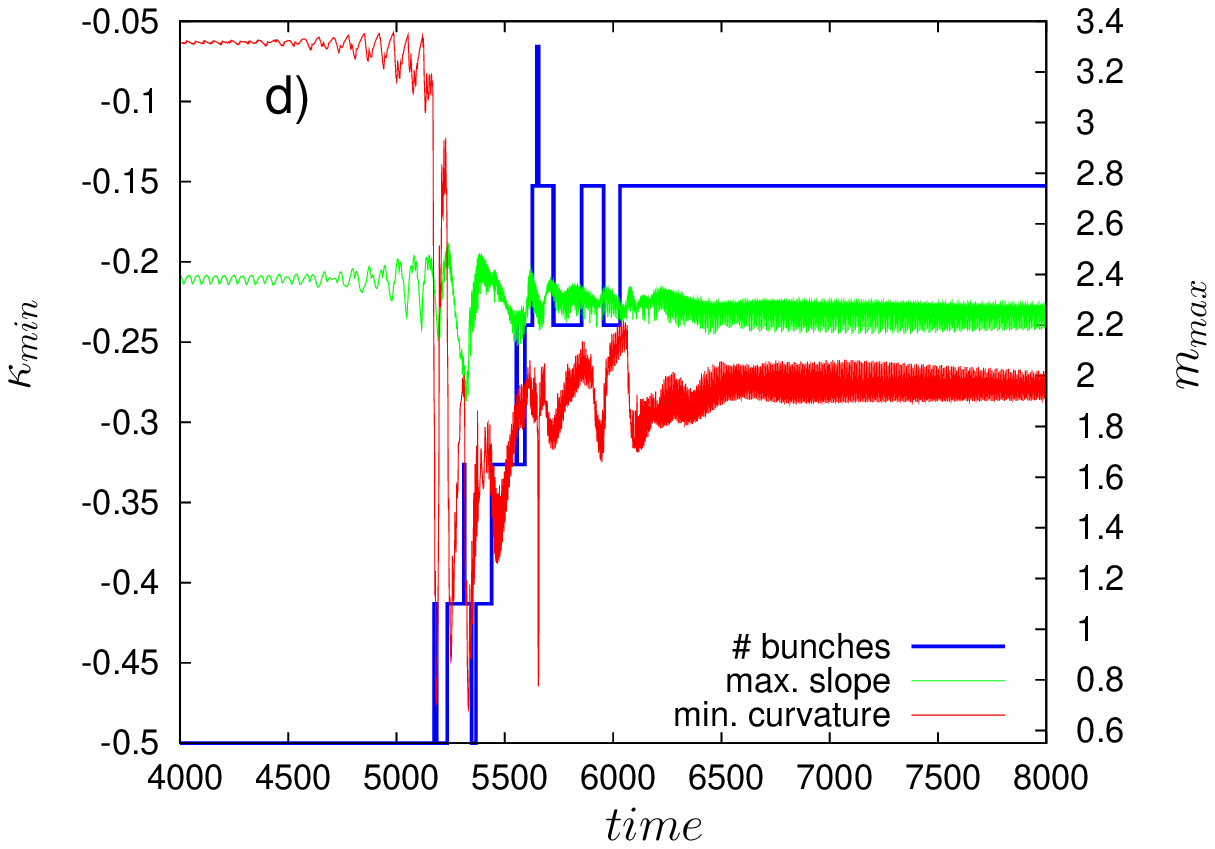}}
\end{center}
\caption{An example for the splitting of a large bunch in a system of
  80 steps with parameters $b=0.7,\  g=0.05,\  U=0.05$. a) Time
  evolution of one of the steps. b) Plot of all step trajectories
  between $5200$t.u. and $5300$t.u. c) Comparison of the profiles
  after 4000 t.u., 6000 t.u., 7000 t.u. and 8000 t.u. d) (Color
  online) Time
  evolution of the globally maximal slope, the globally minimal curvature and the number of bunches.}%
\label{Fig_splitting}%
\end{figure}

The first surprising result we observed is the splitting of a bunch
into two or more bunches after starting with an initial shock of
steps. Let us consider the example in Fig. \ref{Fig_splitting}. The
system consists of 80 steps with parameters  $b=0.7,\  g=0.05$ and$\
U=0.05$. Figure \ref{Fig_splitting}a) shows the time evolution of the
position $x_1$ of one of the steps. There is a constant shift until
the bunch splits, and then the shift increases, which corresponds to a
much faster moving step. In Fig. \ref{Fig_splitting}b) we plot the
positions of all steps around the time of the spliting, where we can
see the emergence of an additional, very small but growing bunch. The
splitting event is followed by another one and so forth until there
are five almost equally sized bunches at 8000 t.u., see the surface
height profiles in \ref{Fig_splitting}c). In
Fig. \ref{Fig_splitting}d) we show the time evolution of the minimal
curvature, the maximal slope and the number of bunches. In the
splitting region the minimal curvature is seen to show a strong
signature with an abrupt jump followed by relaxation to a smaller value,
whereas the change in the maximal slope is rather small. 
The splitting of bunches in this system is an example of
anti-coarsening, where a large structure breaks up into multiple small structures.

\subsection{\label{Sec:arrested}Arrested coarsening}

\begin{figure}
\begin{center}
\scalebox{0.55}{\includegraphics{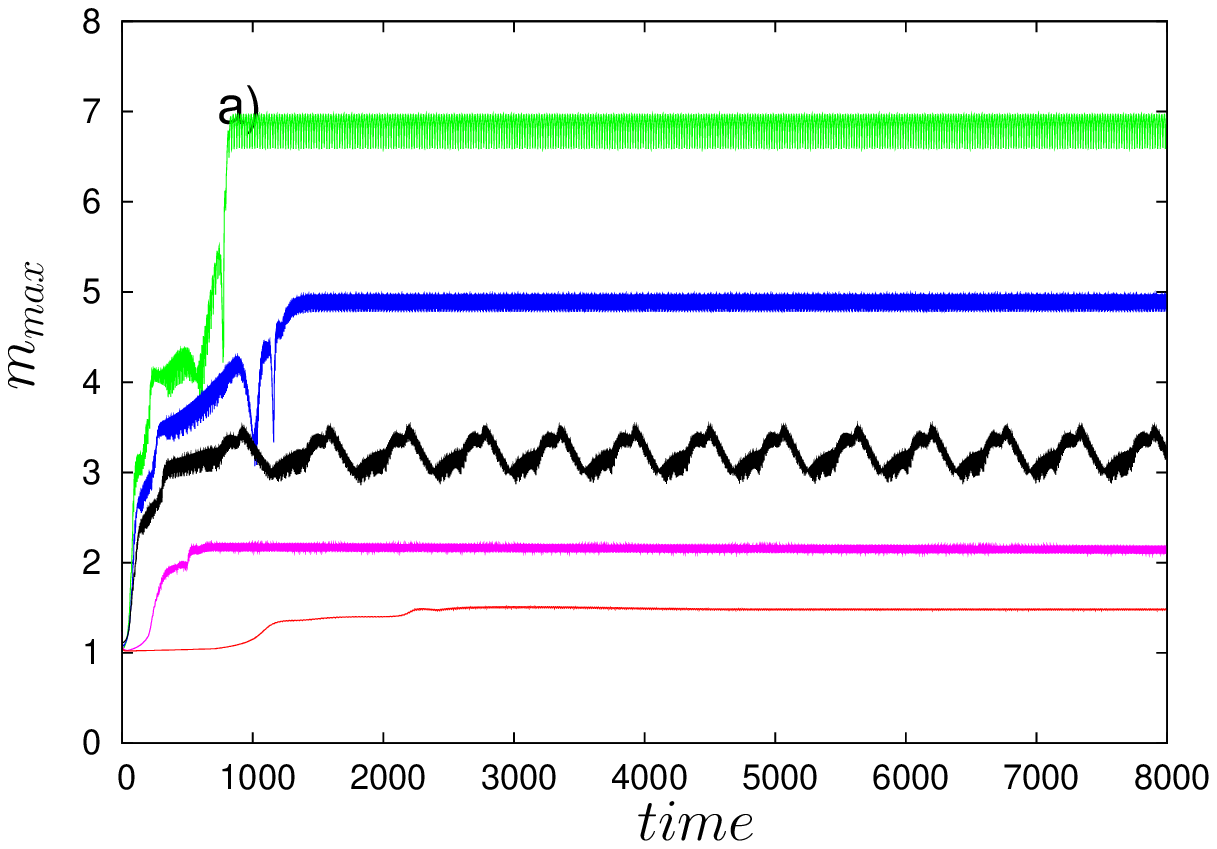}}
\scalebox{0.55}{\includegraphics{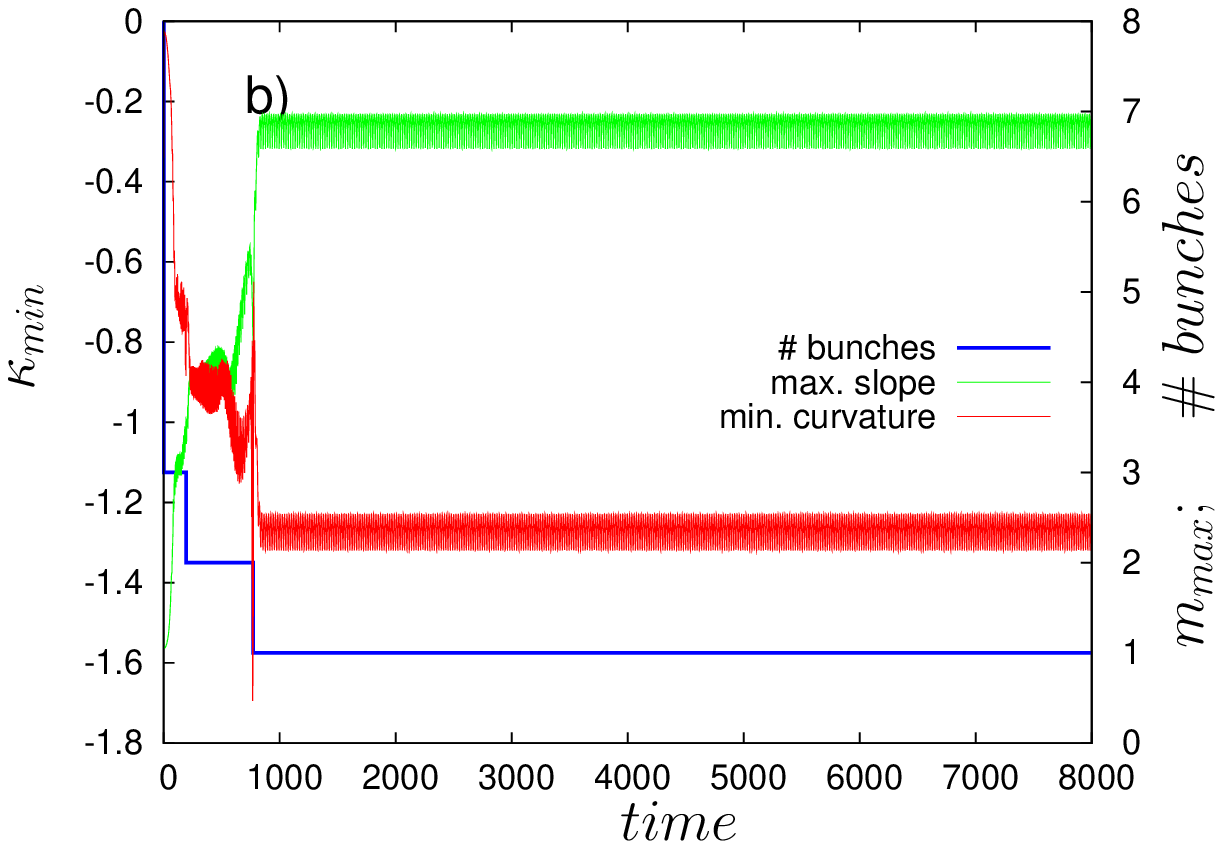}}
\scalebox{0.55}{\includegraphics{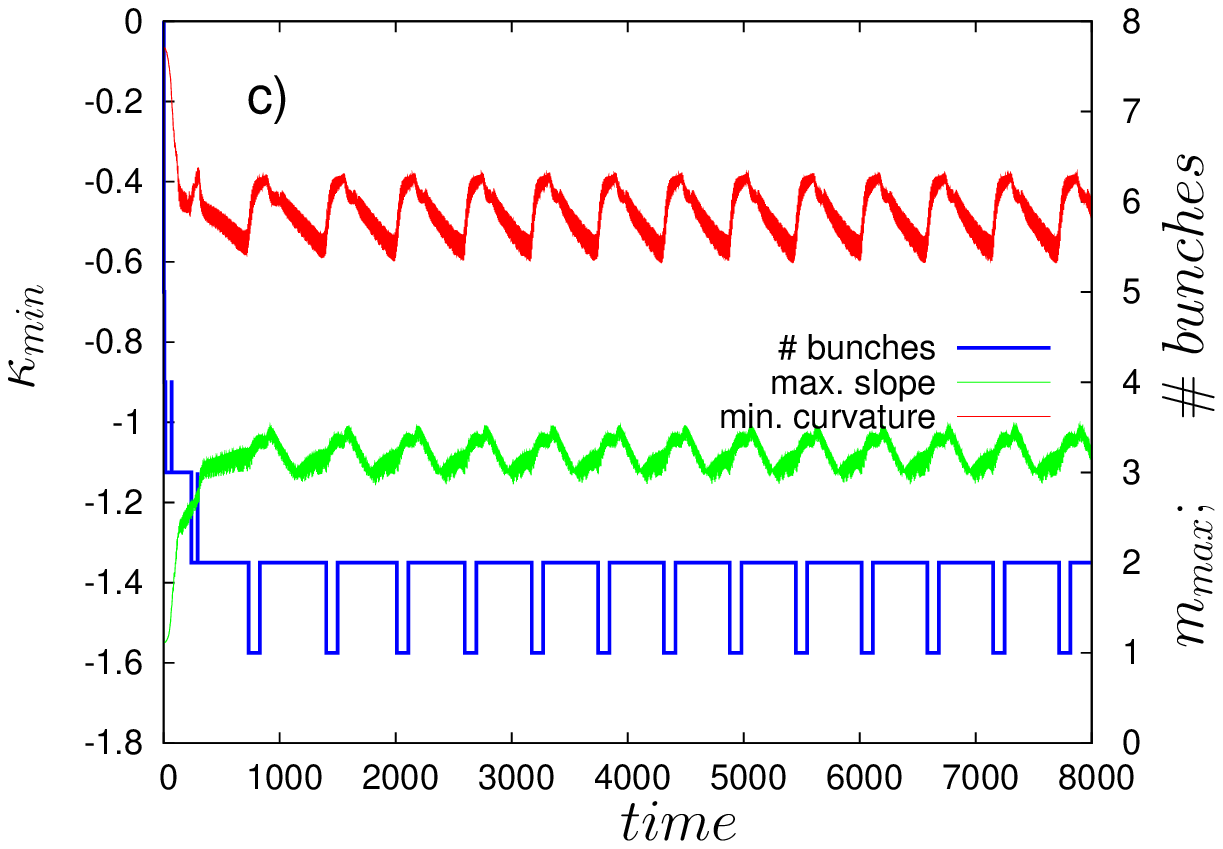}}
\scalebox{0.55}{\includegraphics{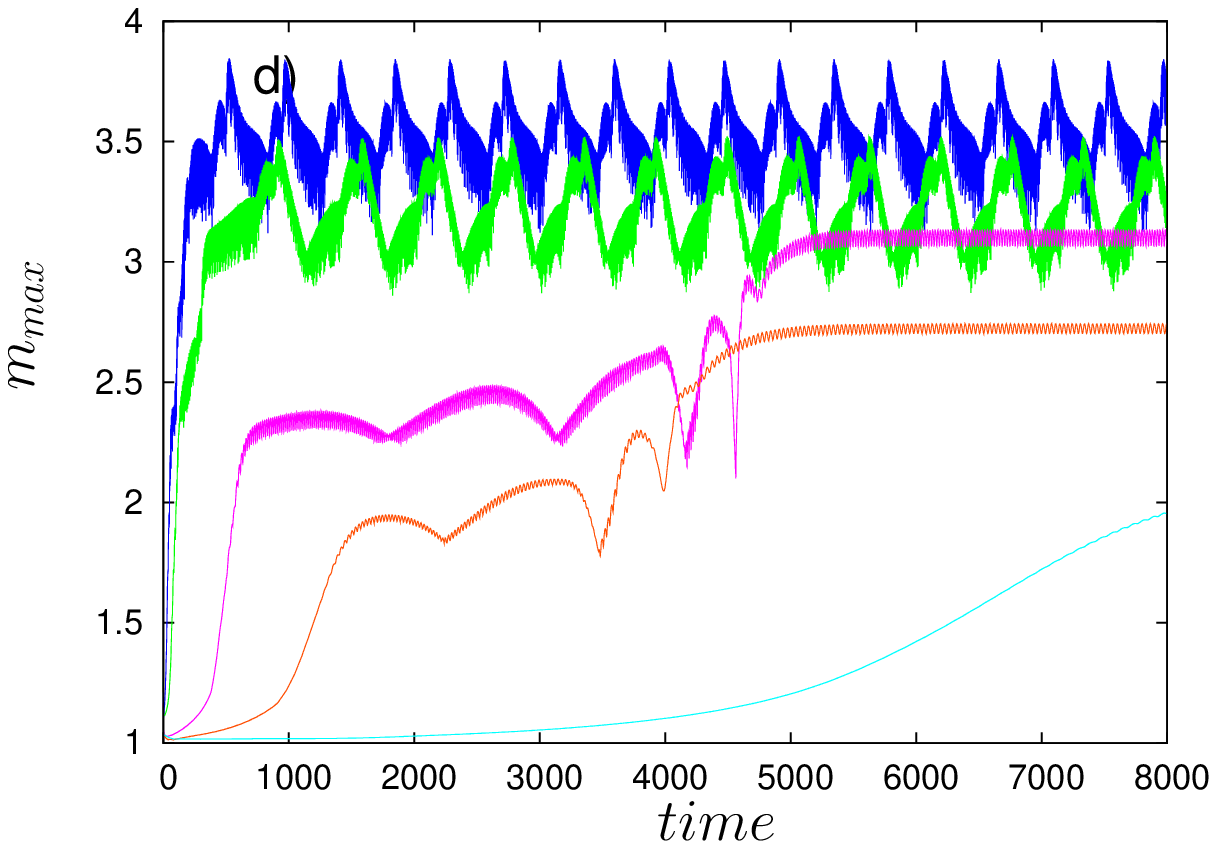}} 
\end{center}
\caption{(Color online) An example of arrested coarsening starting with fluctuating
  initial conditions of strong amplitude, for $M=40$ and $U=0.2$.  a)
  Maximal slope for $b=0.7$ and (from top to bottom) $g=0,\  0.01,\
  0.02,\  0.05,\  0.09$. b) Number of bunches, maximal slope and
  minimal curvature for $b=0.7$ and $g=0$. c) As b) for $b=0.7$ and
  $g=0.02$; d) Maximal slope for $g=0.02$ and (from top to bottom) $b= 0.9,\  0.7,\  0.4,\  0.3,\ 0.2$.}
\label{Fig_g_change}
\end{figure}

The second type of initial condition is a randomly perturbed
equidistant step train. As an example, see Fig. \ref{Fig_g_change}, we
take $M=40$, $U=0.2$, a strong perturbation amplitude, an
integration time of 8000 t.u., and vary the parameters $g$ and $b$. We
use the time evolution of the global maximal slope $m_{max}(t)$ in
order to identify the final state of the simulation. As can be seen in
Fig. \ref{Fig_g_change}a), after about 2000 t.u. the simulations
settled down into a periodic attractor of varying complexity. In
Fig. \ref{Fig_g_change}b) we show the case of vanishing $g$, which
corresponds to a conventional coarsening behavior: After a short time
the number of bunches goes to unity and $m_{max}$ and $\kappa_{min}$
vary periodically on the characteristic time scale in which the bunch
profile shifts by one average terrace size \cite{popkovkrug}. By
increasing $g$ a point is reached where coarsening into a single bunch
no longer occurs, see Fig. \ref{Fig_g_change}c). The system is now
jumping periodically between configurations with two bunches and one
bunch. Upon further increasing $g$ the period of the jumps diverges
and the stable state becomes a configuration of two bunches. 
Finally, Fig. \ref{Fig_g_change}d) displays the various regimes of
temporal behavior that occur when varying $b$ at fixed $g$. 
 \begin{figure}
\begin{center}
\scalebox{0.60}{\includegraphics{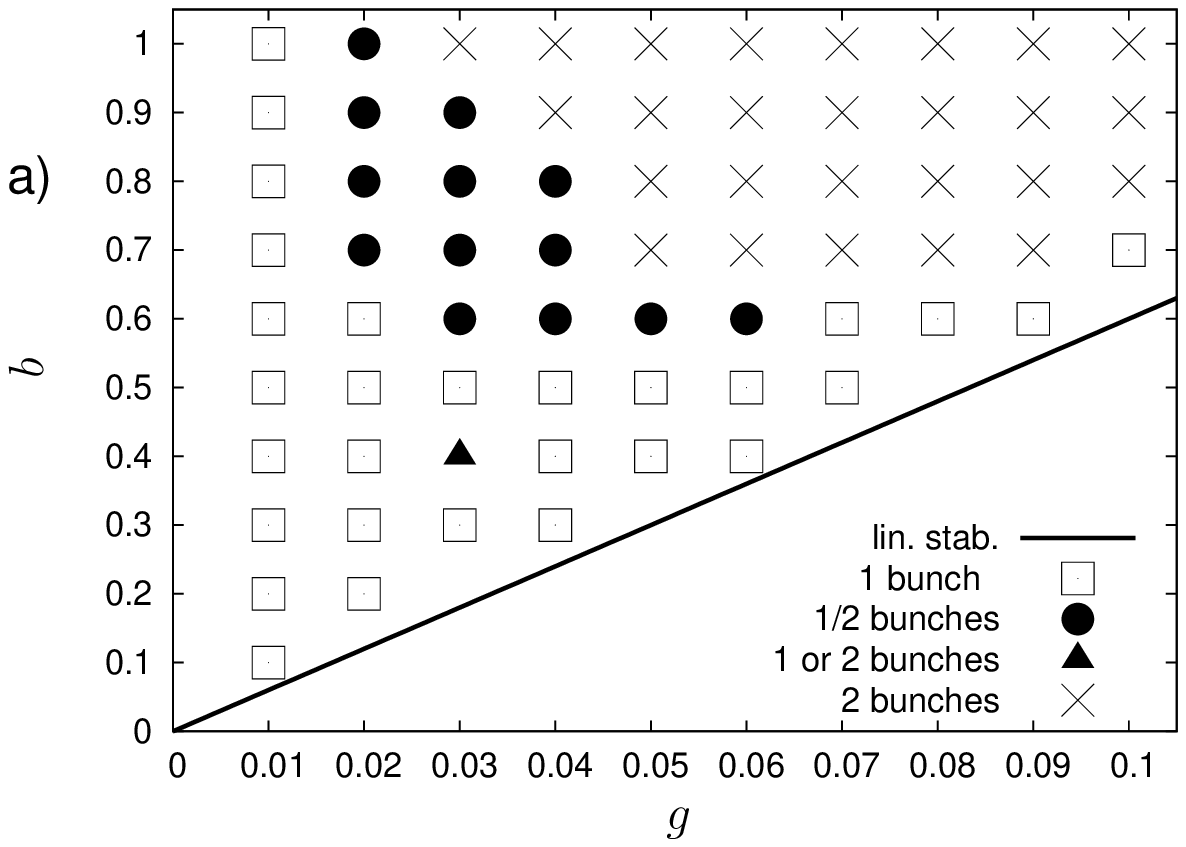}}
\scalebox{0.60}{\includegraphics{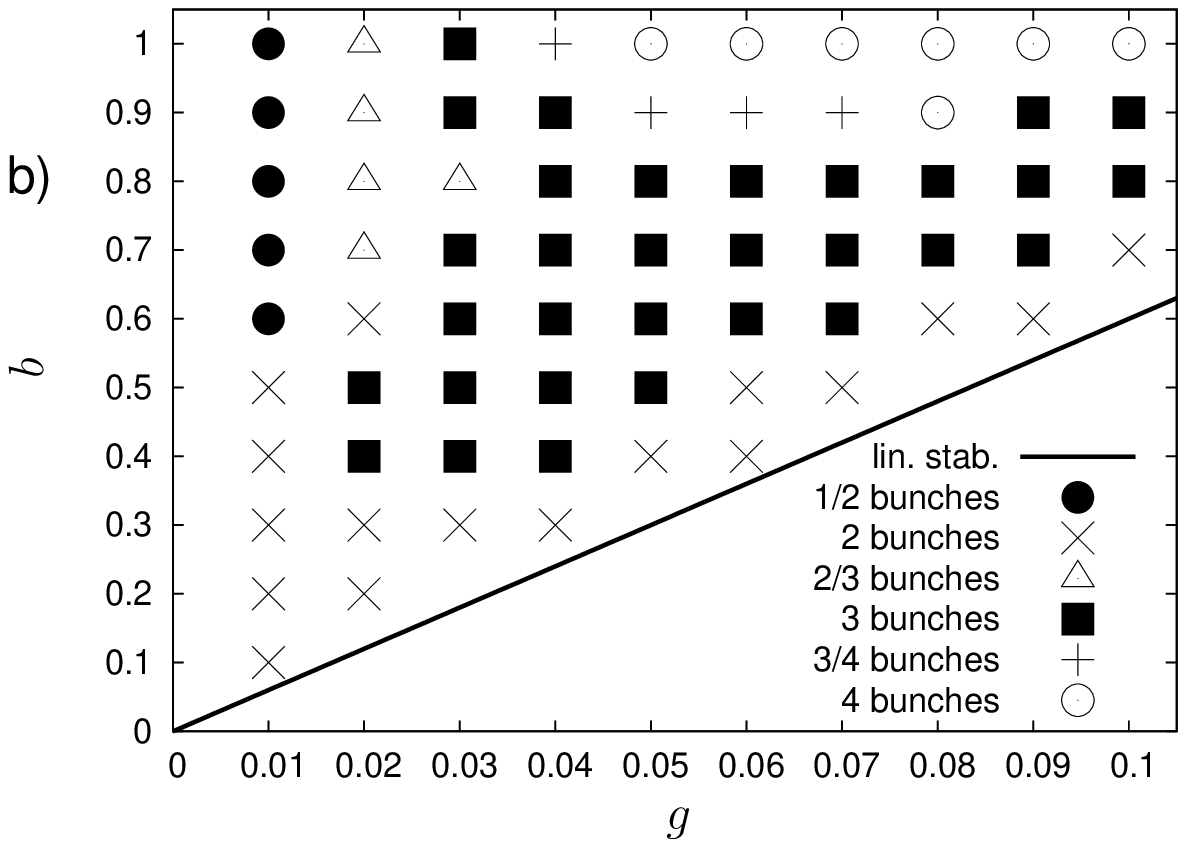}}
\end{center}
\caption{Stability/instability diagrams showing the number of bunches
  in the final state,
  for $M=40$ and different combinations of $b$ and $g$. a) $U=0.2$; b) $U=0.05$  - $\blacktriangle$: region depending sensitive on the initial condition, $\square$: 1 bunch, $\bullet$: 1/2 bunches, $\times$: 2 bunches, $\triangle$: 2/3 bunches, $\blacksquare$: 3 bunches, $+$: 3/4 bunches, $\circ$: 4 bunches, below the line $b=6g$: stability}
\label{Fig_U_change}
\end{figure}
 
\subsection{\label{Sec:diagrams}Phase diagrams}

To somewhat systemize these observations, 
in Fig. \ref{Fig_U_change} we present two phase diagram where the number of
bunches is shown in the $(g,b)$-plane for $b\in[0,1]$ with increments of $\triangle b=0.1$
and $g\in[0,0.1]$ with increments of $\triangle g=0.01$. For $U = 0.2$
[Fig. \ref{Fig_U_change}a)] the number of bunches in the final
stationary state is between one and two. 
At $g=0.02$, $b=0.4$ the relaxation time is very sensitive to the
initial random configuration and one needs more than 8000 t.u. to
switch to the single bunch configuration; moreover, around this point
the final state depends sensitively on initial conditions. For
$U=0.05$ the number of bunches changes between 4 and an oscillatory
state with 1-2 bunches, but a stable state with a single bunch is not
seen. In general, we observe that the smaller $U$ the larger is the
possible number of bunches. For example for $U=0$, $M=40$, $g=0.02$
and $b=0.5$, there are 18 bunches in the final state, corresponding to
less than 2.5 steps pro bunch. 
 
\begin{figure}[ptb]
\scalebox{0.70}{\includegraphics{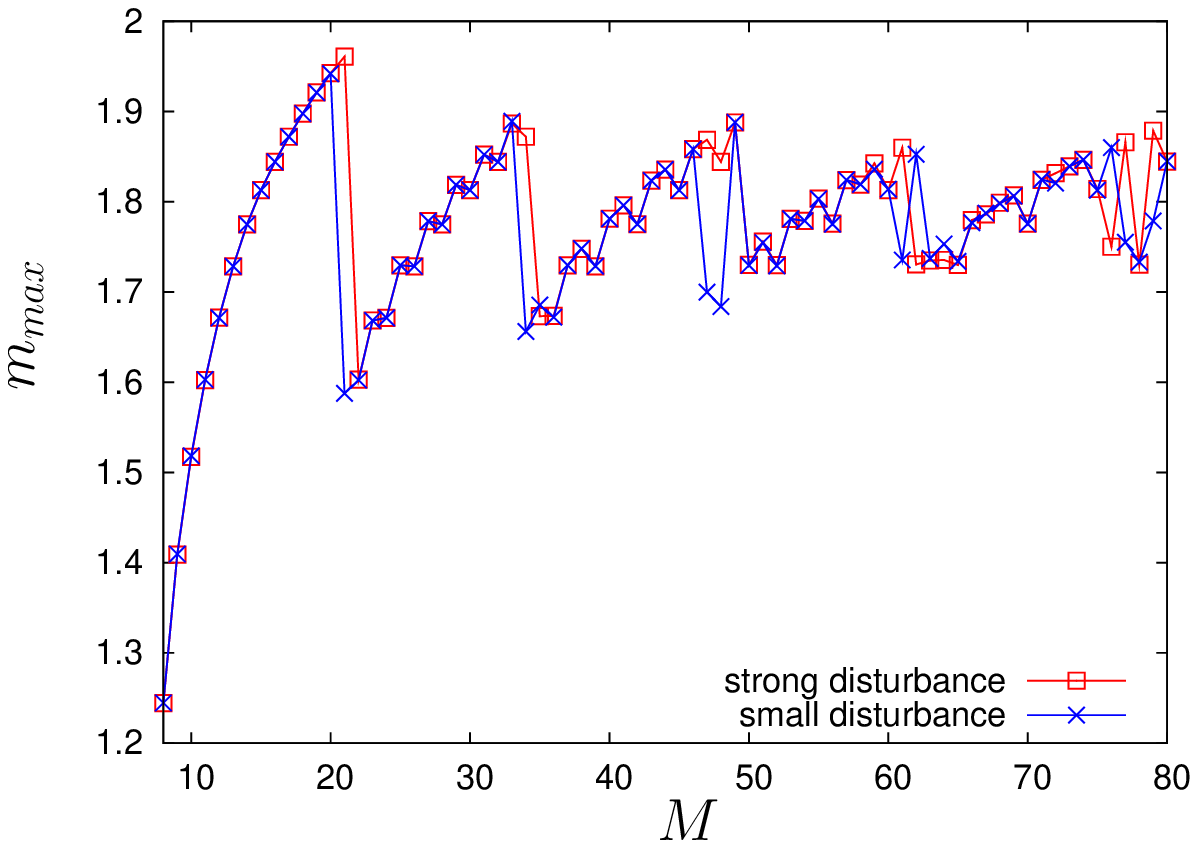}}
\caption{(Color online) Dependence of the upper boundary of the globally maximal slope $m_{max}$ on the number of steps $M$ for $g=U=0.04$, $b=0.4$ and for large and small amplitude of the initial random condition.}%
\label{Fig_M_change}
\end{figure}

In Fig. \ref{Fig_M_change} we show the dependence of the globally
maximal slope $m_{max}$ on $M$ for $g=U=0.04$ and $b=0.4$. As we have
seen in Fig.\ref{Fig_g_change}, $m_{max}$ generally oscillates in the
final state. In order to uniquely define $m_{max}$ we take
for reference its largest value during the last 500 t.u. of the
simulation. For a strong initial disturbance $m_{max}$ grows up to
1.96 at $M=21$ steps and then jumps down to 1.60 at $M=22$, where the
step train switches from a single bunch to a configuration with two
bunches; for a weak initial disturbance the behavior is similar but the
jump occurs slightly earlier. With increasing $M$ this behavior
repeats several times until in the region of 80 steps the switching
becomes less regular and depends strongly on the initial disturbance. 
The overall pattern in Fig.\ref{Fig_M_change} suggests that the
splitting of bunches is driven by the tendency of the system to keep
the maximal slope below a certain limiting value, thus connecting the
behavior of the discrete system to the properties of stationary
solutions of the continuum equation discussed in Sect.\ref{Sec:Cont}.

\section{\label{Sec:Summary}Summary and discussion}

In this paper we have investigated the extension
(\ref{eq:discretemodel}) of a previously studied, minimal model of
step bunching during sublimation.
Despite the smallness of the dimensionless coefficient $g$ describing
the interplay between sublimation and capillarity, the
non-conservative nature of the dynamics was found to have profound
effects on the linear as well as on the nonlinear level. In order to
analytically study the nonlinear behavior of step bunches, we derived
and analyzed the continuum evolution equation (\ref{Continuous}) for
the surface profile. Using two different types of approximations, we
deduced that the non-conservative terms in this equation place an
upper limit on the slope as well as on the wavelength of step
bunches. This conclusion was confirmed by a detailed numerical study
of the discrete step dynamics, which also revealed additional complex
dynamical regimes in which the number of step bunches increases in
time or oscillates between different values.

To place our findings into context, we note that a connection between   
unbounded coarsening of step bunches and the conservative nature of
the dynamics has been observed in previous work. First, Sato and Uwaha 
studied step bunching induced by electromigration and
observed a saturation of the mean bunch size when sublimation was
included \cite{Sato1999}. A direct comparison to our results is
difficult, however, since they use a different set of step dynamical
equations which depend on a larger set of parameters. 

Second, the difference between conserved and non-conserved surface
dynamics has been addressed in the framework of weakly nonlinear 
(in the sense of \cite{PierreLouis2005b}) continuum equations. 
Working close to the instability threshold, Sato and Uwaha
\cite{Sato1995} and Misbah and Pierre-Louis \cite{Misbah1996} showed
that the large-scale dynamics of step bunches in a non-conserved
setting is described by the Benney equation. Depending on 
the size of a term that breaks the left-right symmetry, this equation
displays either spatio-temporal chaos or an ordered array of bunches,
but no coarsening. On the other hand, the corresponding conservative
equation derived in \cite{Gillet2001} does show unlimited coarsening
with the average bunch size growing as $t^{1/2}$. In general, deducing the coarsening
behavior of a given nonlinear evolution equation  by analytic means is
a difficult problem, and the methods available so far
\cite{Politi2006}
do not seem to readily carry over to highly nonlinear equations like
(\ref{Continuous}).     

\section*{Acknowledgements} We thank P. Politi,
B. Ranguelov, S. Stoyanov and V. Tonchev for useful discussions.

\begin{appendix}

\section{Discrete equations for the case of sublimation}
\subsection{\label{app:profile}The concentration profile}
In the quasistatic approximation $\partial n_i/\partial t = 0$
\eqref{eq:BCFmodel} with $F = 0$ is a homogeneous ordinary differential equation of second order with the following general solution:
\begin{equation}\label{eq:app1a}
n_i(x)=C_1^ie^{\frac{x}{l_D}}+C_2^ie^{-\frac{x}{l_D}}.
\end{equation}
In order to find the constants of integration $C_1^i$ and $C_2^i$ we use the boundary conditions \eqref{eq:bc}:
\begin{eqnarray}\label{eq:app1b}\nonumber
\frac{D_s}{k_-}\left (\frac{C_1^i}{l_D} e^{-\frac{l_i}{2l_D}} -\frac{C_2^i}{l_D}e^{+ \frac{l_i}{2l_D}}  \right )&=&+C_1^ie^{-\frac{l_i}{2l_D}} +C_2^ie^{+ \frac{l_i}{2l_D}}  -n^0_{eq}\left (1+\frac{\mu_i}{kT}\right ),\\ \nonumber
\frac{D_s}{k_+}\left (\frac{C_1^i}{l_D} e^{+\frac{l_i}{2l_D}} -\frac{C_2^i}{l_D}e^{- \frac{l_i}{2l_D}}  \right )&=&-C_1^ie^{+\frac{l_i}{2l_D}} -C_2^ie^{-\frac{l_i}{2l_D}}  +n^0_{eq}\left (1+\frac{\mu_{i+1}}{kT}\right ).
\end{eqnarray}
Substituting $l^{\pm}:=\frac{D_s}{l_Dk_{\pm}}=\frac{l_{\pm}}{l_D}$
this becomes
\begin{eqnarray}\label{eq:app1b2}\nonumber
 (l^--1)C_1^ie^{-\frac{l_i}{2l_D}} -(l^-+1)C_2^ie^{+ \frac{l_i}{2l_D}}  &=& -n^0_{eq}\left (1+\frac{\mu_i}{kT}\right ),\\ \nonumber
 (l^++1)C_1^ie^{+\frac{l_i}{2l_D}} -(l^+-1)C_2^ie^{- \frac{l_i}{2l_D}}  &=& +n^0_{eq}\left (1+\frac{\mu_{i+1}}{kT}\right ).
\end{eqnarray}
After adding both equations and solving for $C_1^i$ we obtain
\begin{eqnarray}\label{eq:app1c}
C_1^i=\frac{n^0_{eq}\left (\frac{\mu_{i+1}}{kT}-\frac{\mu_{i}}{kT}\right )+C_2^i\left [ \left (l^-+1 \right ) e^{\frac{l_i}{2l_D}}+\left (l^+-1 \right ) e^{-\frac{l_i}{2l_D}}\right ]}{\left (l^--1 \right ) e^{-\frac{l_i}{2l_D}}+\left (l^++1 \right ) e^{\frac{l_i}{2l_D}}}.
\end{eqnarray}
We substitute $C_1^i$ in the first boundary condition:
\begin{eqnarray}\label{eq:app1d}
C_2^i=\frac{n^0_{eq}\left (1+\frac{\mu_{i}}{kT}\right )+C_1^i(l^--1)e^{-\lambda_1\frac{l_i}{2}}}{(l^-+1)e^{-\lambda_2\frac{l_i}{2}}}.
\end{eqnarray}
After back substitution from  $C_2^i$ into $C_1^i$ and vice versa we get the final form for the constants: 
\begin{eqnarray}\label{eq:app1e}\nonumber
C_1^i&=&\frac{n^0_{eq}}{2}\left [ \frac{(l^+-1)(1+\frac{\mu_i}{kT})e^{-\frac{l_i}{2l_D}}+(l^-+1)(1+\frac{\mu_{i+1}}{kT})e^{\frac{l_i}{2l_D}}}{(l^+l^-+1)\sinh\frac{l_i}{l_D}+(l^-+l^+)\cosh\frac{l_i}{l_D}}\right ],\\
C_2^i&=&\frac{n^0_{eq}}{2}\left [\frac{(l^--1)(1+\frac{\mu_i}{kT})e^{-\frac{l_i}{2l_D}}+(l^++1)(1+\frac{\mu_{i+1}}{kT})e^{\frac{l_i}{2l_D}}}{(l^+l^-+1)\sinh\frac{l_i}{l_D}+(l^-+l^+)\cosh\frac{l_i}{l_D}}\right ].
\end{eqnarray}

\subsection{\label{app:fluxes}Fluxes and the step velocities}
Using the definition
\begin{equation*}\label{eq:app1f}
f^i_{\pm}(x)=\pm\Omega D_s\frac{\partial n_i(x)}{\partial x}  \  \  \text{at}\ x=\pm\frac{l_i}{2}
\end{equation*}
we find for the fluxes to the step bordering the $i$th terrace of width $l_i$:
\begin{eqnarray}\label{eq:app1g}\nonumber
f_+^i  &=& \frac{\Omega D_sn^0_{eq}}{l_D}\frac{\left (1+\frac{\mu_{i+1}}{kT}\right )\left (l^-\sinh\frac{l_i}{l_D}+\cosh\frac{l_i}{l_D}\right )-\left (1+\frac{\mu_{i}}{kT}\right )}{(l^+l^-+1)\sinh\frac{l_i}{l_D}+(l^-+l^+)\cosh\frac{l_i}{l_D}}, \  \  \text{at step}\  i+1, \\  \nonumber
f_-^i &=&\frac{\Omega D_sn^0_{eq}}{l_D}\frac{\left (1+\frac{\mu_{i}}{kT}\right )\left (l^+\sinh\frac{l_i}{l_D}+\cosh\frac{l_i}{l_D}\right )-\left (1+\frac{\mu_{i+1}}{kT}\right )}{(l^+l^-+1)\sinh\frac{l_i}{l_D}+(l^-+l^+)\cosh\frac{l_i}{l_D}}, \  \  \text{at step}\  i.
\end{eqnarray}
Adding the fluxes from the two neighboring terraces we obtain the
expression for the velocity of step $i$ as  $v_i=\frac{dx_i}{dt}=f_-^{i}+f_+^{i-1}$.

\subsection{\label{app:linear}Linear stability analysis}
Let us assume that $l_D\gg l$ and $l_D\gg l_{\pm}$. Then $\sinh\frac{l_i}{l_D}\approx \frac{l_i}{l_D}$ and $\cosh\frac{l_i}{l_D}\approx1.$
For the denominator in the expression for the fluxes (and with
$l^{\pm}=\frac{l_{\pm}}{l_D}$) we find
\begin{eqnarray}\label{eq:app1i}
(\frac{l_+l_-}{l_D^2}+1)\sinh\frac{l_{i}}{l_D}+(\frac{l_-}{l_D}+\frac{l_+}{l_D})\cosh\frac{l_{i}}{l_D}&\approx&\frac{l_-+l_++l_{i}}{l_D},
\end{eqnarray}
and the velocity becomes
\begin{eqnarray}\label{eq:app1j}\nonumber
\frac{1}{\Omega D_s n^0_{eq}}\frac{dx_i}{dt} &\approx&\frac{\left (1+\frac{\mu_{i}}{kT}\right )\left (\frac{l_-l_{i-1}}{l^2_D}+1\right )-\left (1+\frac{\mu_{i-1}}{kT}\right )}{l_-+l_++l_{i-1}} + \frac{\left (1+\frac{\mu_{i}}{kT}\right )\left (\frac{l_+l_{i}}{l^2_D}+1\right )-\left (1+\frac{\mu_{i+1}}{kT}\right )}{l_-+l_++l_{i}} \\ 
&=&\frac{\left (1+\frac{\mu_{i}}{kT}\right )\frac{l_-l_{i-1}}{l^2_D}+(\frac{\mu_{i}}{kT}-\frac{\mu_{i-1}}{kT})}{l_-+l_++l_{i-1}} + \frac{\left (1+\frac{\mu_{i}}{kT}\right )\frac{l_+l_{i}}{l^2_D}+(\frac{\mu_{i}}{kT}-\frac{\mu_{i+1}}{kT})}{l_-+l_++l_{i}} 
\end{eqnarray}
The chemical potential $\mu_{i}$ can be linearized as 
\begin{eqnarray}\label{eq:app1k}\nonumber
\frac{\mu_{i}}{kT} \approx - g\left [1-3\frac{\varepsilon_{i+1}-\varepsilon_i}{l}-1+3\frac{\varepsilon_{i}-\varepsilon_{i-1}}{l} \right ]=-\frac{3g}{l}(2 \varepsilon_i -\varepsilon_{i+1} -\varepsilon_{i-1}) \Rightarrow
\end{eqnarray}
\begin{eqnarray}\label{eq:app1l}\nonumber
 \frac{\mu_{i}}{kT}-\frac{\mu_{i\pm 1}}{kT} &\approx& -\frac{3g}{l}(-3 \varepsilon_{i\pm1}+3\varepsilon_{i} +\varepsilon_{i\pm2} -\varepsilon_{i\mp1})\\ \nonumber
\left (1+\frac{\mu_{i}}{kT}\right )\frac{l_{\pm}l_{i,i-1}}{l^2_D} &\approx&
\frac{l_{\pm}}{l^2_D}(l+\varepsilon_{i+1,i} -\varepsilon_{i,i-1})-\frac{3gl_{\pm}}{l^2_D}(2 \varepsilon_i -\varepsilon_{i+1}-\varepsilon_{i-1})
\end{eqnarray}
\begin{eqnarray}\label{eq:app1m}\nonumber
 \frac{1}{\Omega D_s n^0_{eq}}\frac{dx_i}{dt}&\approx& \frac{(l_{+}+l_{-})l}{(l_-+l_++l)l^2_D}+\frac{3g}{(l_-+l_++l)l}(-6\varepsilon_{i}+4\varepsilon_{i+1}+4\varepsilon_{i-1}-\varepsilon_{i+2}-\varepsilon_{i-2})\\ \nonumber 
&+&\frac{l_{-}(\varepsilon_{i}-\varepsilon_{i-1})+l_{+}(\varepsilon_{i+1}-\varepsilon_{i})}{(l_-+l_++l)l^2_D}
-\frac{3g(l_++l_-)}{(l_-+l_++l)l^2_D}(2\varepsilon_{i}-\varepsilon_{i+1}+\varepsilon_{i-1})\\
&\doteq&\frac{1}{\Omega Dn^0_{eq}} \left(v_{eq}+\frac{d\varepsilon_i}{dt} \right)
\end{eqnarray}
The equation for the perturbation reads
\begin{eqnarray}\label{eq:app1n}\nonumber
\frac{1}{\Omega D_s n^0_{eq}} \frac{d\varepsilon_i}{dt}&=&\frac{3g}{(l_-+l_++l)l}(-6\varepsilon_{i}+4\varepsilon_{i+1}+4\varepsilon_{i-1}-\varepsilon_{i+2}-\varepsilon_{i-2})\\ \nonumber 
&+&\frac{l_{-}(\varepsilon_{i}-\varepsilon_{i-1})+l_{+}(\varepsilon_{i+1}-\varepsilon_{i})}{(l_-+l_++l)l^2_D}
-\frac{3g(l_++l_-)}{(l_-+l_++l)l^2_D}(2\varepsilon_{i}-\varepsilon_{i+1}+\varepsilon_{i-1})
\end{eqnarray}
Using the Fourier expression $\varepsilon=\varepsilon_0e^{ikn+\omega(k)t}$ we find the dispersion relation $\omega(k)$:
\begin{eqnarray}\label{eq:app1o} \nonumber
\frac{\omega(k)}{\Omega D_s n^0_{eq}} &=& \frac{3g}{(l_-+l_++l)l}[-6+4(e^{ik}+e^{-ik})-(e^{i2k}+e^{-i2k})] \\ \nonumber
&+&\frac{l_{-}(1-e^{-ik})+l_{+}(e^{ik}-1)}{(l_-+l_++l)l^2_D}
-\frac{3g(l_++l_-)}{(l_-+l_++l)l^2_D}(2-(e^{ik}+e^{-ik})).
\end{eqnarray}
For small $k$ we can expand the exponential functions as
\begin{eqnarray}\label{eq:app1p}\nonumber
2-(e^{ik}+e^{-ik})&=&2(1-\cos k)\approx2(1-1+\frac{k^2}{2})=k^2,\\\nonumber
6-4(e^{ik}+e^{-ik})+(e^{i2k}+e^{-i2k})&=&4(\cos k-1)^2\approx k^4,\\ \nonumber
1-e^{-ik}&\approx&1-(1-ik+\frac{k^2}{2})=ik+\frac{k^2}{2},
\end{eqnarray}
which yields an expression for the real part $Re[\omega]$:
\begin{eqnarray}\label{eq:app1q}\nonumber
Re[\omega]=\frac{\Omega D_s n^0_{eq}}{l^2_D}\frac{(l_-+l_+)}{(l_-+l_++l)} \left [ \frac{(l_--l_+)}{2} -3g\right] k^2-\frac{\Omega D_s n^0_{eq}3g}{(l_-+l_++l)l}k^4.
\end{eqnarray}
Further, if $l_{\pm}\gg l$
\begin{eqnarray}\label{eq:app1r}
Re[\omega]=\frac{\Omega D_s n^0_{eq}}{l^2_D}\left [ \frac{(k_+-k_-)}{2(k_++k_-)} -3g\right] k^2-3g\frac{\Omega D_s n^0_{eq}}{l}\frac{k_-k_+}{k_-+k_+}k^4.
\end{eqnarray}
For instability $Re[\omega]$ has to be positive. The prefactor $A_4$
of the $k^4$-term is always negative and thus acts as a relaxation
term. For $k$ very small the linear instability condition $b>6g$ ($A_2>0$) follows.
\subsection{\label{app:limit}Discrete equations in the limit $l_D\gg l_{\pm} \gg l_i$}
From equation \eqref{eq:app1j} with $l_{\pm} \gg l_i$ we get
\begin{eqnarray}\label{eq:app1s}\nonumber
\frac{dx_i}{dt} &\approx&\frac{\Omega D_s n^0_{eq}}{l_-+l_+} \left (1+\frac{\mu_{i}}{kT}\right )\left(\frac{l_-l_{i-1}}{l^2_D}+\frac{l_+l_{i}}{l^2_D}\right)+\frac{\Omega D_s n^0_{eq}}{l_-+l_+}\left(2\frac{\mu_{i}}{kT}-\frac{\mu_{i-1}}{kT}-\frac{\mu_{i+1}}{kT}\right)\\ \nonumber
&=&(1+g\nu_i)R_e \left (\frac{l_-l_{i-1}+l_+l_{i}}{l_-+l_+} \right )+UR_e(2\nu_i-\nu_{i-1}-\nu_{i+1})\\
&=&\left (1+g\nu_i\right )R_e \left (  \frac{1+b}{2}l_{i-1}+\frac{1-b}{2}l_{i}\right )+ UR_e(2\nu_i-\nu_{i-1}-\nu_{i+i})
\end{eqnarray}

\section{Continuum limit}
\label{app:cont}

Similar to \cite{JoachimKim}, we treat the part of (\ref{eq:discretemodel}) without
the $U$-term first,%

\begin{equation}
\frac{dx_{n}}{dt}=\left(  1+g\nu_{n}\right)  \left(  \frac{1-b}{2}l
_{n}+\frac{1+b}{2}l_{n-1}\right)  \label{eq_linearpart}%
\end{equation}
where $l_{n}=x_{n+1}-x_{n}$.
To obtain the continuum limit of \ $\left(  \frac{1-b}{2}l_{n}+\frac{1+b}%
{2}l_{n-1}\right)$, we rewrite it in the form
\begin{align*}
\frac{dx_{n}}{dt}  &  =\left(  1+g\nu_{n}\right)  \left(  \frac{l
_{n}+l_{n-1}}{2}-\frac{b}{2}\left(  l_{n}-l_{n-1}\right)
\right) \\
\frac{dx_{n+1}}{dt}  &  =\left(  1+g\nu_{n+1}\right)  \left(  \frac{l
_{n+1}+l_{n}}{2}-\frac{b}{2}\left(  l_{n+1}-l_{n}\right)
\right)
\end{align*}
Substracting,
\begin{align*}
\frac{dl_{n}}{dt}  &  =\frac{l_{n+1}-l_{n-1}}{2}-\frac{b}%
{2}\left(  l_{n+1}-2l_{n}+l_{n-1}\right) \\
&  +g\frac{\nu_{n+1}l_{n+1}-\nu_{n}l_{n}}{2}+g\frac{\nu_{n+1}l
_{n}-\nu_{n}l_{n-1}}{2}-gb\frac{\nu_{n+1}l_{n+1}+\nu_{n}l
_{n-1}-\left(  \nu_{n+1}+\nu_{n}\right)  l_{n}}{2}%
\end{align*}
Let us treat the $g$-independent term first. Performing the Fourier transform
$l_{n}=\sum_{q}e^{iqn}l_{q}$, we obtain
\[
\sum_{q}e^{iqn}\frac{dl_{q}}{dt}=\sum_{q}e^{iqn}\left[  \frac{1}%
{2}\left(  e^{iq}-e^{-iq}\right)  -b\frac{e^{iq}+e^{-iq}-2}{2}\right]
l_{q}%
\]
Expanding for small $q$,%
\[
\sum_{q}e^{iqn}\frac{dl_{q}}{dt}\approx\sum_{q}e^{iqn}\left[  \left(
(iq)+\frac{(iq)^{3}}{3!}+...\right)  -b\left(  \frac{(iq)^{2}}{2!}%
+\frac{(iq)^{4}}{4!}\right)  \right]  l_{q}%
\]
Noting that $(iq)\sum_{q}e^{iqn}l_{q}=\frac{\partial}{\partial n}\left(
\sum_{q}e^{iqn}l_{q}\right)  =\frac{\partial}{\partial n}l_{n}$
etc., one can rewrite the previous equation as
\[
\frac{dl_{n}}{dt}=\left[  \left(  \frac{\partial}{\partial n}+\frac
{1}{3!}\frac{\partial^{3}}{\partial n^{3}}\right)  -b\left(  \frac{1}{2!}%
\frac{\partial^{2}}{\partial n^{2}}+\frac{1}{4!}\frac{\partial^{4}}{\partial
n^{4}}\right)  \right]  l_{n}.%
\]
Now, we can use $\frac{\partial}{\partial n}=h_{0}\frac{\partial}{\partial
h}=\frac{\partial}{\partial h}$ (note that $h_{0}=1$) and $l
_{n}\approx1/m(x)=1/(\partial h/\partial x)=\partial x/\partial h$. We obtain
\[
\frac{\partial}{\partial h}\frac{dx}{dt}=\frac{\partial}{\partial h}\left[
\left(  1+\frac{1}{3!}\frac{\partial^{2}}{\partial h^{2}}\right)  -b\left(
\frac{1}{2!}\frac{\partial}{\partial h}+\frac{1}{4!}\frac{\partial^{3}%
}{\partial h^{3}}\right)  \right]  \Delta\left(  x\right)
\]
where $\Delta(x) = 1/m(x)$. 
Integrating this equation we find
\[
\frac{dx}{dt}=\left[  \left(  1+\frac{1}{3!}\frac{\partial^{2}}{\partial
h^{2}}\right)  -b\left(  \frac{1}{2!}\frac{\partial}{\partial h}+\frac{1}%
{4!}\frac{\partial^{3}}{\partial h^{3}}\right)  \right]  \frac{\partial
x\left(  t\right)  }{\partial h}.
\]
Inserting the terms proportional to $g$ and approximating
\begin{equation}
\nu_{n}=\left(  \frac{l}{l_{n-1}}\right)  ^{3}-\left(  \frac{l}{l_{n}%
}\right)  ^{3}\approx-l_{n}\frac{\partial}{\partial x}\left(  \frac
{l}{\Delta}\right)  ^{3}=-\frac{1}{m}\left(  m^{3}\right)  ^{\prime}=-\frac
{3}{2}\left(  m^{2}\right)  ^{\prime} \label{fn}%
\end{equation}
we get%
\[
\frac{dx}{dt}=\left(  1-\frac{3g}{2}\left(  m^{2}\right)  ^{\prime}\right)
\left[  \left(  1+\frac{1}{3!}\frac{\partial^{2}}{\partial h^{2}}\right)
-b\left(  \frac{1}{2!}\frac{\partial}{\partial h}+\frac{1}{4!}\frac
{\partial^{3}}{\partial h^{3}}\right)  \right]  \Delta\left(  x,t\right)
\]

The next important step is to carry out the Lagrange transformation
\cite{JoachimKim,Krug1997b} from $x(n,t)=x(h=nh_{0},t)=x(h,t)$ to $h(x,t)$. First, note that $\frac{dx}{dt}=-\left(  \frac
{dx}{dh}\right)  \frac{dh}{dt}=-\Delta\frac{dh}{dt}$. Substituting this and
using $\frac{\partial}{\partial h}=\left(  \frac{dx}{dh}\right)
\frac{\partial}{\partial x}=\Delta\frac{\partial}{\partial x}$,
\[
-\frac{dh}{dt}=\left(  1-\frac{3g}{2}\left(  m^{2}\right)  ^{\prime}\right)
+\left(  1-\frac{3g}{2}\left(  m^{2}\right)  ^{\prime}\right)  \frac{\partial
}{\partial x}\left[  \left(  \frac{1}{3!}\frac{\partial}{\partial h}\right)
-b\left(  \frac{1}{2!}+\frac{1}{4!}\frac{\partial^{2}}{\partial h^{2}}\right)
\right]  \Delta\left(  x,t\right)
\]
or
\[
-1=\frac{dh}{dt}+\frac{\partial}{\partial x}\left\{  -\frac{3g}{2}\left(
m^{2}\right)  +\left[  \left(  \frac{1}{3!}\frac{\partial \Delta}{\partial
h}\right)  -b\left(  \frac{\Delta}{2!}+\frac{1}{4!}\frac{\partial^{2}\Delta
}{\partial h^{2}}\right)  \right]  \right\} 
\] 
\[ -\frac{3g}{2}\left(
m^{2}\right)  ^{\prime}\frac{\partial}{\partial x}\left[  \left(  \frac{1}%
{3!}\frac{\partial \Delta}{\partial h}\right)  -b\left(  \frac{\Delta}%
{2!}+\frac{1}{4!}\frac{\partial^{2}\Delta}{\partial h^{2}}\right)  \right]
\]
Now we estimate the term in the square brackets, using $\Delta=1/m$%
,$\frac{\partial}{\partial h}=\Delta\frac{\partial}{\partial x}=(1/m)\frac
{\partial}{\partial x}$
\[
\left[  \left(  \frac{1}{3!}\frac{\partial \Delta}{\partial h}\right)
-b\left(  \frac{\Delta}{2!}+\frac{1}{4!}\frac{\partial^{2}\Delta}{\partial
h^{2}}\right)  \right]  =\frac{1}{6m}\frac{\partial}{\partial x}\left(
\frac{1}{m}\right)  -\frac{b}{2m}-\frac{b}{24}\frac{\partial^{2}\Delta
}{\partial h^{2}}%
\]
Aiming at smooth solutions, we ignore high order derivatives, approximating
\[
\left[  ...\right]  \approx-\frac{b}{2m}-\frac{m^{\prime}}{6m^{3}}.
\]
Substituting we obtain
\[
\frac{dh}{dt}+\frac{\partial}{\partial x}\left(  -\frac{3g}{2}\left(
m^{2}\right)  -\frac{b}{2m}-\frac{m^{\prime}}{6m^{3}}\right)  =-1+\frac{3g}%
{2}\left(  m^{2}\right)  ^{\prime}\left[  -\frac{b}{2m}-\frac{m^{\prime}%
}{6m^{3}}\right]  ^{\prime}%
\]
The term proportional to $U$ gives a contribution to the full derivative,%
\[
\frac{dh}{dt}+\frac{\partial}{\partial x}\left(  -\frac{3g}{2}\left(
m^{2}\right)  -\frac{b}{2m}-\frac{m^{\prime}}{6m^{3}}+\frac{3U}{2}%
\frac{\left(  m^{2}\right)  ^{\prime\prime}}{m}\right)  =-1+\frac{3g}%
{2}\left(  m^{2}\right)  ^{\prime}\left[  -\frac{b}{2m}-\frac{m^{\prime}%
}{6m^{3}}\right]  ^{\prime}%
\]
The terms proportional to $g$ on the RHS break the conservative character of the
equation, and are important for the dynamics as discussed in the main text.
However, our numerical data (not shown) suggest that the best (though not
perfect) agreement with the discrete system is obtained by keeping only the
nonconservative term of the lowest order,%
\[
\frac{dh}{dt}+\frac{\partial}{\partial x}\left(  -\frac{3g}{2}\left(
m^{2}\right)  -\frac{b}{2m}-\frac{m^{\prime}}{6m^{3}}+\frac{3U}{2}%
\frac{\left(  m^{2}\right)  ^{\prime\prime}}{m}\right)  =-1+\frac{3g}%
{2}\left(  m^{2}\right)  ^{\prime}\left[  -\frac{b}{2m}\right]  ^{\prime}%
\]
Expanding the derivative on the RHS, we obtain (\ref{Continuous}).
\end{appendix}


\end{document}